\documentclass{article}
\pdfoutput=1
\usepackage[utf8]{inputenc}
\usepackage{amsmath}
\usepackage[a4paper, total={6in, 8in}]{geometry}
\usepackage{graphicx}
\usepackage{color}

\usepackage{amssymb}
\usepackage[sorting=none]{biblatex}
\usepackage[T1]{fontenc}
\usepackage[utf8]{inputenc}
\usepackage{authblk}
\usepackage{subfig} 
\usepackage{graphicx}

\title{\textbf{Voter-like model with conflicting preferences on homophilic networks}}
\title{\textbf{Voter-like dynamics with conflicting preferences on modular networks}}
\vspace{5cm}
\author{Filippo Zimmaro$^{1,2,3}$, Pierluigi Contucci$^2$, János Kertész$^3$}
\date{%
    $^1$ \textit{Department of Computer Science}, University of Pisa\\%
    $^2$ \textit{Department of Mathematics}, University of Bologna \\%
    $^3$ \textit{Department of Network and Data Science}, Central European University\\[4ex]%
}

\begin{document}
\maketitle
\begin{abstract} Two of the main factors shaping an individual's opinion are social coordination and personal preferences, or personal biases. To understand the role of those and that of the topology of the network of interactions, we study an extension of the voter model proposed by Masuda and Redner (2011), where the agents are divided into two populations with opposite preferences.
We consider a modular graph with two communities that reflect the bias assignment, modeling the phenomenon of epistemic bubbles. We analyze the models by approximate analytical methods and by simulations. Depending on the network and the biases' strengths, the system can either reach a consensus or a polarized state, in which the two populations stabilize to different average opinions. The modular structure has generally the effect of increasing both the degree of 
polarization and its range in the space of parameters. 
When the difference in the bias strengths between the populations is large, the success of the very committed group in imposing its preferred opinion onto the other one depends largely on the level of segregation of the latter population, while the dependency on the topological structure of the former is negligible. We compare the simple mean-field approach with the pair approximation and test the goodness of the mean-field predictions on a real network.
\end{abstract}
\section{Introduction}
The formation of people's opinions, choices and decisions is subject to social pressure: it is a general observation that in society, individuals (agents) take into account the behaviour of others \cite{weber1978economy}. This aspect is at the root of most agent-based models of opinion dynamics \cite{castellano2009statistical,sirbu2017opinion,contucci2020statistical,peralta2022opinion}. Many such models have been introduced with the aim to understand the effects of 
different microscopic mechanisms of the opinion formation process \cite{redner2019reality,castello2006ordering}. Here, we consider the Voter model \cite{holley1975ergodic,dornic2001critical,sood2005voter}, characterized by a simple imitative mechanism, introducing personal preferences attached to single individuals \cite{masuda2010heterogeneous,masuda2011can}. \\
\\
Prejudices or personal preferences generally come from the history of the individual (e.g. ideologies and partisanship \cite{bartels2000partisanship}) and are assumed to evolve on a much longer time-scale than that of the opinion influencing interactions, in other words they 
can be considered as fixed (quenched) features of the nodes throughout the dynamics\footnote{Recent studies \cite{alford2005political,settle2009heritability,settle2010friendships} corroborate this assumption by providing evidence of a genetic contribution in the formation of
political attitudes, exhorting scientists to \textit{"incorporate genetic influences, specifically interactions between genetic heritability and social environment, into models of political attitude formation.}" \cite{alford2005political}.}. Even though this bias is a characteristic of the node, 
it is fundamentally different from other kinds of biases, such as the confirmation or algorithmic ones \cite{sirbu2019algorithmic,peralta2021effect,peralta2021opinion}, since those are dependent on the node's current opinion.\\
\\
In a model with social pressure and quenched preferences, each individual is subjected to two "forces": one inducing the individual to minimize conflicts with his neighbours, the other exhorting the individual to stick to his own prejudice.
A strong social pressure may lead the agent to adopt a public opinion in dissonance with his prejudices, a phenomenon which is defined as \textit{preference falsification} in \cite{kuran1997private}. Else, if the personal bias is strong, the individual may reject social coordination, accept conflicts more easily, and stick to his prior view even if he finds himself in disagreement with many of his neighbours. The latter mechanism, when conflicting preferences are present among the individuals, may contribute
to the emergence of \textit{polarization} \cite{valensise2022dynamics,del2017modeling}.\\
\\
In this work, we generally consider two groups of interacting individuals with opposite preferences of different intensities. We focus in particular on the role of the social network, considering a modular network with two communities corresponding to the bias assignment. The network model 
mimics the realistic setting of two \textit{epistemic bubbles}\footnote{We remark, following the subtle distinction highlighted by \cite{nguyen2020echo}, that the \textit{epistemic bubble} is a structure related to the process of acquiring knowledge that emerges principally due to an "unconscious" homophilic mechanism, and it is considerably different from the \textit{echo chamber}, which involves more active processes, e.g. the mechanisms of \textit{trust-distrust}, \textit{isolation} and \textit{disagreement-reinforcement}.}, where the agents with the same ideology share more links among themselves than with the other oppositely biased community. The result is an unbalanced choice of sources by the agent, who is systematically more influenced by his
own community than by the other. In this paper, our main focus resides in determining the level of polarization between the two groups once the system has reached the stationary state, as a function of the preferences' intensities and of the topological structure of the social network \cite{interian2022network,garimella2018political}. \\
\\
There is an analogy between the binary opinion dynamics models and the Ising model of statistical physics, where the personal preference can be achieved by a site-dependent external magnetic field. Following this analogy, we will characterize the opinion state of the individuals by a $\sigma \in \{+1,-1\}$ "spin" variable. We study the so-called \textit{Partisanship Voter Model} (PVM), in which the preference toward one of the states modifies the transition rates of the voter dynamics accordingly and breaks the original symmetry between the two opinions. This dynamics should be distinguished from that of another biased voter model that we call \textit{Voter model with media interactions} (VMMI) \cite{bhat2020polarization}, where the personal bias expresses how much an individuum follows a node with fixed opinion state connected to everybody (the "medium"). In table \ref{table:single} 
we summarize the transition rates of the Ising model and the biased voter models for homogeneous populations, where all individuals have the same personal biases. The same models in the bipopulated versions \cite{gallo2007bipartite,contucci2008phase} are defined and explained in table \ref{table:bipopulated}.\\
\\
The PVM with homogeneous preferences can be historically individuated as a specific case of the Abrams-Strogatz model for language death \cite{abrams2003modelling} and as an agent-based model in \cite{stauffer2007microscopic, vazquez2010agent}. The generalization to multiple biases has been first proposed by Masuda et al. in 2010 \cite{masuda2010heterogeneous}. The authors of \cite{borile2013effect} focused in particular on the finite-size effects for low bias intensity.  In the successive work \cite{masuda2011can}, the model was generalized for different compositions of the system and preferences' intensities. In \cite{czaplicka2022biased} the same model was considered where just a fraction of individuals was biased. Let us also point out that the introduction of \textit{zealots} \cite{mobilia2003does,mobilia2007role,mukhopadhyay2020voter}, i.e., agents that never change opinion and just try to convince others, can be traced back to the analyzed model if we properly tune the bias associated to such agents (i.e. setting $h_{z} = \{+1,-1\}$ depending on the type of commitment). The problem of social pressure and conflicting preferences has been also 
studied in evolutionary game theory \cite{hernandez2013heterogeneous,hernandez2017equilibrium,mazzoli2017equilibria}, with a focus on network effects \cite{broere2017network} and supported with various social experiments \cite{ellwardt2016conflict,goyal2021integration,broere2019experimental}. \\
\\
Despite the advancements in understanding the dynamics of systems evolving with Voter-like processes, the influence of the social network topology in the PVM with conflicting preferences remains unexplored. This paper aims to address this gap by examining the interplay between fixed individual preferences and homophilic network structures (epistemic bubbles). Indeed, we expect homophily to play a crucial role in mitigating the social cohesion induced by imitative dynamics, when conflicting preferences are present, possibly leading the system to a polarized asymptotic state.\\  
\\
In this work, we consider the model of Masuda and Redner in its most general version \cite{masuda2011can}, that we refer to as (Bipopulated) \textit{Voter Model with Preferences} (VMP) \footnote{Since personal preferences can arise from various factors, including partisanship, we use this term in the most general sense to connect our model to existing literature, such as game-theoretical models.}, where we consider two classes of agents with opposite biases of different intensities. 
The VMP is defined in section \ref{model} and solved on the fully connected network in section \ref{FC_chapter}. In section \ref{chapter SBM}, we study the model on a bi-modular network and calculate the phase diagram as a function of the model parameters using a mean-field approach.
In section \ref{PA_chapter}, we apply the Pair Approximation \cite{gleeson2011high,gleeson2013binary} to the model on the modular network, and compare its predictions to the mean-field results for sparse graphs. In the remainder of the article, we study the model on a real network with high modularity, the Political blogosphere of 2004 US elections \cite{adamic2005political}, and test the goodness of the mean-field predictions of the stationary state in the case of equally intense but opposite personal biases.\\

\begin{table}[ht]
\centering
\begin{tabular}{ |p{3cm}|p{2.8cm}|p{2.8cm}|
} 
\hline
 & 
 $F_{k,m}$ & 
 $R_{k,m}$ 
 \\[0.5ex] 
 \hline
 Ising Glauber 
 & $\frac{1}{1+e^{-\frac{2}{T}[h+J(2m-k)]}}$ & $\frac{e^{-\frac{2}{T}[h+J(2m-k)]}}{1+e^{-\frac{2}{T}[h+J(2m-k)]}}$
 \\[0.5ex]
 \hline
 VMMI
 & $(1-h)\frac{m}{k} + h$ & $(1-h)(1-\frac{m}{k})$ 
 \\  [0.5ex] 
 \hline
 PVM& $\frac{m}{k}(\frac{1+h}{2}) $ & $(1-\frac{m}{k})(\frac{1-h}{2})$ 
 \\  [0.5ex] 
 \hline\hline
\end{tabular}
\caption{\textbf{Binary-state models with homogeneous personal biases (preferences).} Ising Glauber: Ising model with Glauber dynamics, VMMI: Voter Model with Media Interaction, PVM: Partisanship Voter Model. The models are defined through the infection and recovery transition rates $F_{k,m}$ and $R_{k,m}$ for $\sigma:-1 \to +1$ and $\sigma:+1 \to -1$, respectively; the considered node has $k$ neighbors out of which $m$ are in state $+1$. The strength of the bias is $h$, which corresponds to the external field in the Ising model with temperature $T$ and pairwise couplings $J$. For the voter models $h\in [0,1].$}
\label{table:single}
\end{table}

\begin{table}[ht]
\centering
\begin{tabular}{ |p{2.1cm}|p{2.4cm}|p{2.4cm}|p{2.4cm}|p{2.7cm}| }
 \hline
 & 
 $F^{(1)}_{k,m}$ & 
 $R^{(1)}_{k,m}$ & 
 $F^{(2)}_{k,m}$ & 
 $R^{(2)}_{k,m}$  \\[0.5ex] 
 \hline
 Ising Glauber 
 & $\frac{1}{1+e^{-\frac{2}{T}[h_1+J(2m-k)]}}$ & $\frac{e^{-\frac{2}{T}[h_1+J(2m-k)]}}{1+e^{-\frac{2}{T}[h_1+J(2m-k)]}}$ &  $\frac{1}{1+e^{-\frac{2}{T}[h_2+J(2m-k)]}}$ & $\frac{e^{-\frac{2}{T}[h_2+J(2m-k)]}}{1+e^{-\frac{2}{T}[h_2+J(2m-k)]}}$ \\[0.5ex]
 \hline
 VMMI
 & $(1-h_1)\frac{m}{k} + h_1$ & $(1-h_1)(1-\frac{m}{k})$ & $(1-|h_2|)\frac{m}{k}$ & $(1-|h_2|)(1-\frac{m}{k})+|h_2|$  \\  [0.5ex] 
 \hline
 PVM & $\frac{m}{k}(\frac{1+h_1}{2}) $ & $(1-\frac{m}{k})(\frac{1-h_1}{2})$ & $\frac{m}{k}(\frac{1+h_2}{2}) $ & $(1-\frac{m}{k})(\frac{1-h_2}{2})$ \\
 \hline\hline
\end{tabular}
\caption{\textbf{Bipopulated binary-state models with personal biases (preferences).} Columns two and three (four and five) are for nodes in population 1 (2). For the VMMI, $h_1\in [0,1]$ and $h_2\in [-1,0]$.}
\label{table:bipopulated}
\end{table}

\section{Voter Model with Preferences} \label{model}
The VMP, i.e. the generalization of the PVM of Masuda et al. \cite{masuda2010heterogeneous, masuda2011can}, is defined as follows: the system of $N$ agents is divided into two populations, or \textit{classes}, of sizes $N_1$ and $N_2 = N - N_1$, the agents $i = 1,...,N_1$ belonging to the first and the remaining $i = N_1+1,...,N$ to the second one, with $\alpha = \frac{N_1}{N}\in(0,1)$ the fraction of individuals of the first population. A bias $h_i\in[-1,1]$ is assigned to each agent $i$, according to his class: in our bipopulated case, we assign the same $h_1$ to the individuals of the first population, similarly $h_2$ to all the individuals belonging to the second one. Each node's opinion is represented as a binary spin $\sigma_i=\{+1,-1\}$ for $i=1,...,N$. The dynamics obeys the following rules:
\begin{itemize}
    \item One node (agent) $i$ is 
    selected uniformly randomly.
    \item A neighbor $j$ of node $i$ is selected uniformly randomly.
    \item If $i$ and $j$ have opposite opinions, $i$ takes the opinion of $j$ with probability $\frac{1}{2}(1+\sigma_jh_i)$. Otherwise, nothing happens.
    \item Repeat the process until consensus or apparent stabilization is reached.
\end{itemize}
The dynamics is a generalization of the classical voter model - retrieved for $h_i=0,\;\forall i$ - where the individual copies his neighbour's opinion with a probability equal to $\frac{1}{2}$ (in the original voter model this probability is 1, but the factor $\frac{1}{2}$ just slows down the dynamics). The biases modify the transition probabilities, favoring the transition towards the direction of the bias and disfavoring the opposite one. It is easy to show that if in the bipopulated case both of the biases point towards the same direction, then an infinite system will always reach consensus at the preferred state. Thus, in the following we will consider $h_1\geq0$ and $h_2\leq0$, so that the individuals of the first population tend to prefer the $+1$ state, while the ones of the second class are ideologically biased towards the $-1$ opinion.

\section{VMP on the complete graph}
\label{FC_chapter}
\begin{figure}
    \subfloat{\includegraphics[width = 3in]{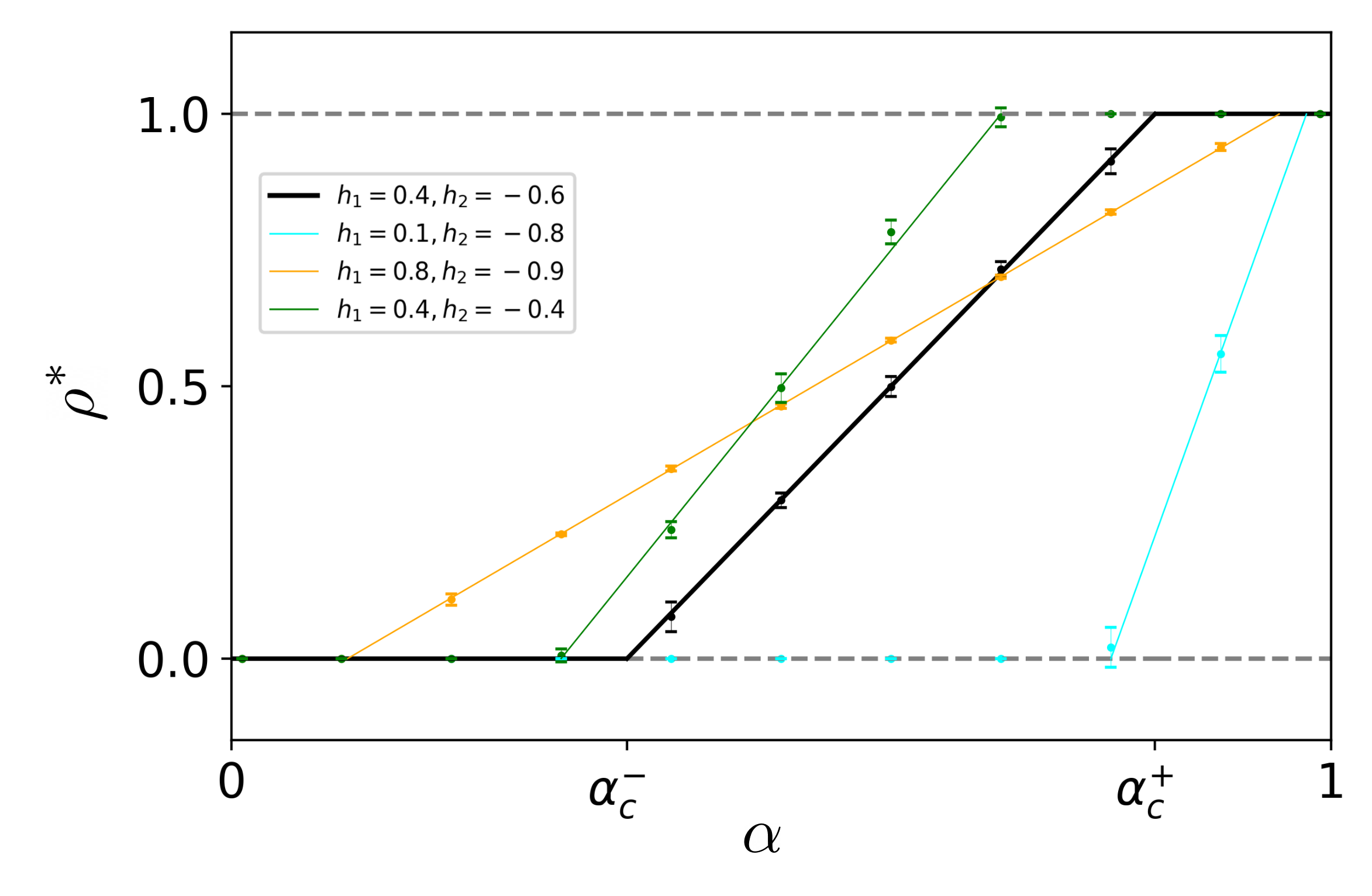}}
    \subfloat{\includegraphics[width = 3in]{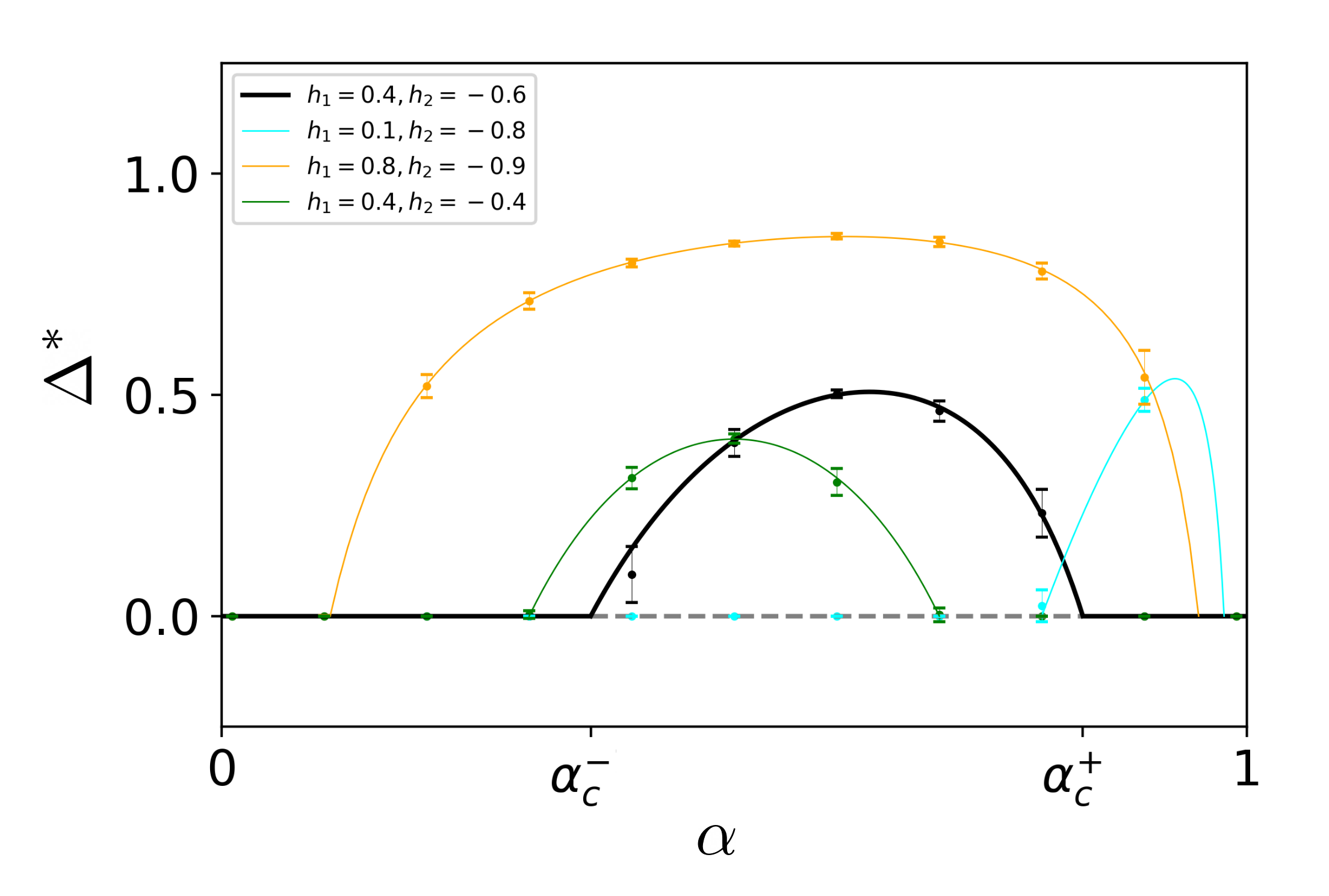}}\\
    \caption{\textbf{Bifurcation diagrams for the bipopulated Voter Model with Preferences on the complete network.} On the right the total density of up spin is taken as order parameter, on the left the polarization $\Delta$ is shown. The solid black line indicates the stable fixed point, while the dashed gray lines indicate the unstable ones, for the choice of the preferences' intensities $h_1=0.4,h_2=-0.6$. The other solid colored lines locate just the coexistence stable fixed point for other choices of the intensities, as indicated in the legend. The points and their bars are respectively the average and the confidence interval of the order parameters calculated over $30$ independent simulations of a system with $1000$ agents, and are reported in order to test the validity of the mean field treatment also for relatively small system sizes.}
    \label{fig:fc_bifurcation}
\end{figure}

\begin{figure}
\subfloat(a){\includegraphics[width = 2.7in]{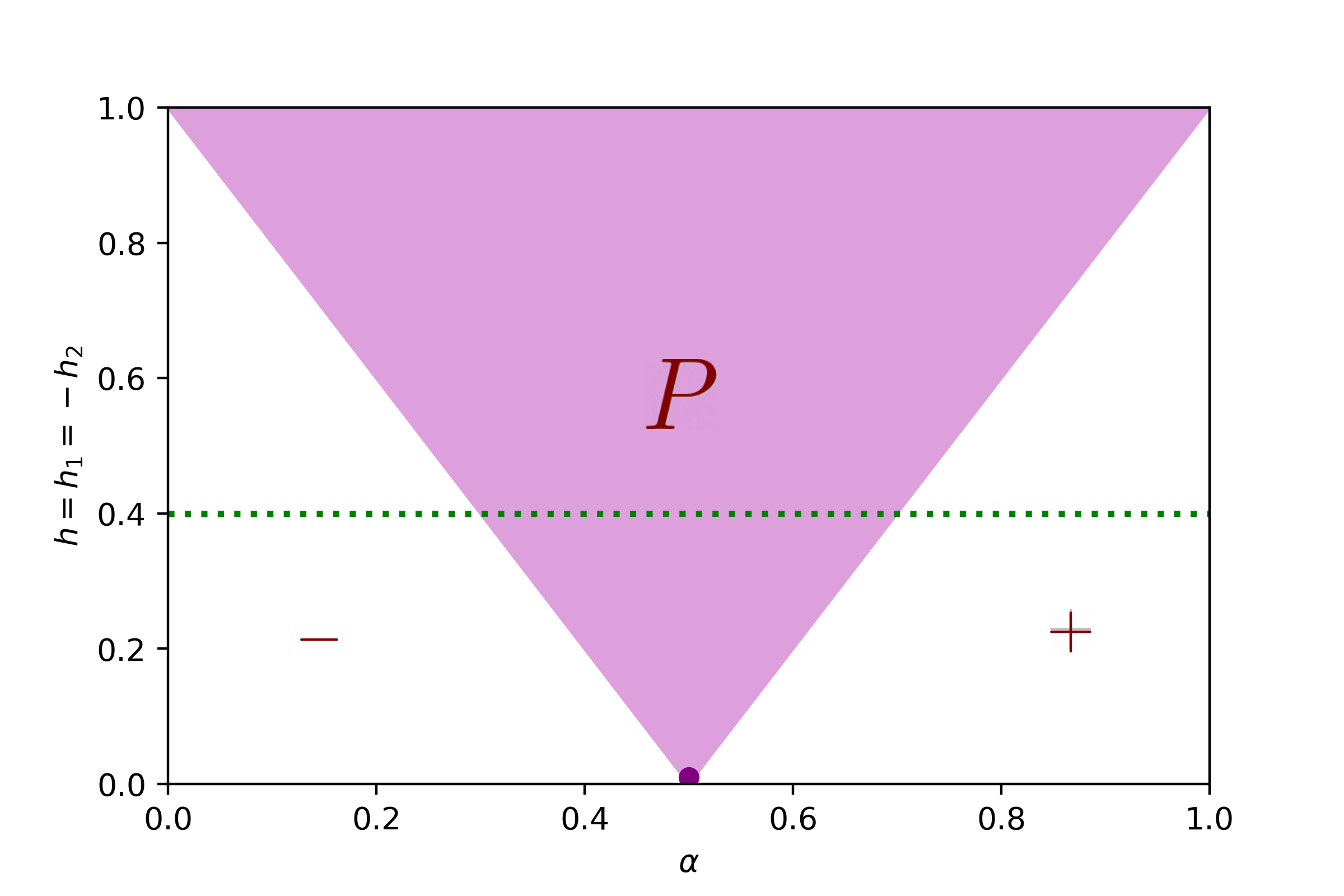}}
\subfloat(b){\includegraphics[width = 2.7in]{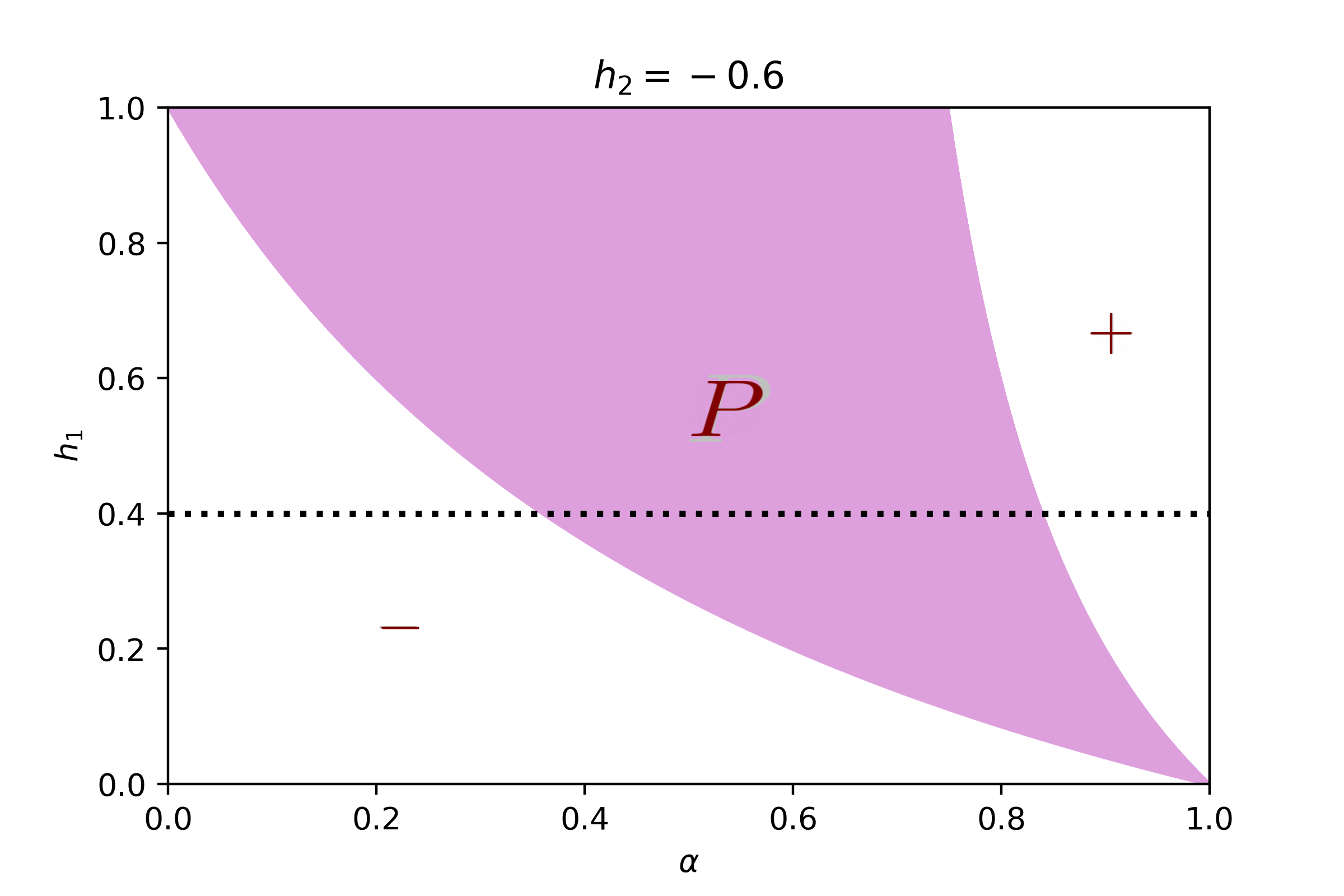}}\\
\subfloat(c){\includegraphics[width = 2.7in]{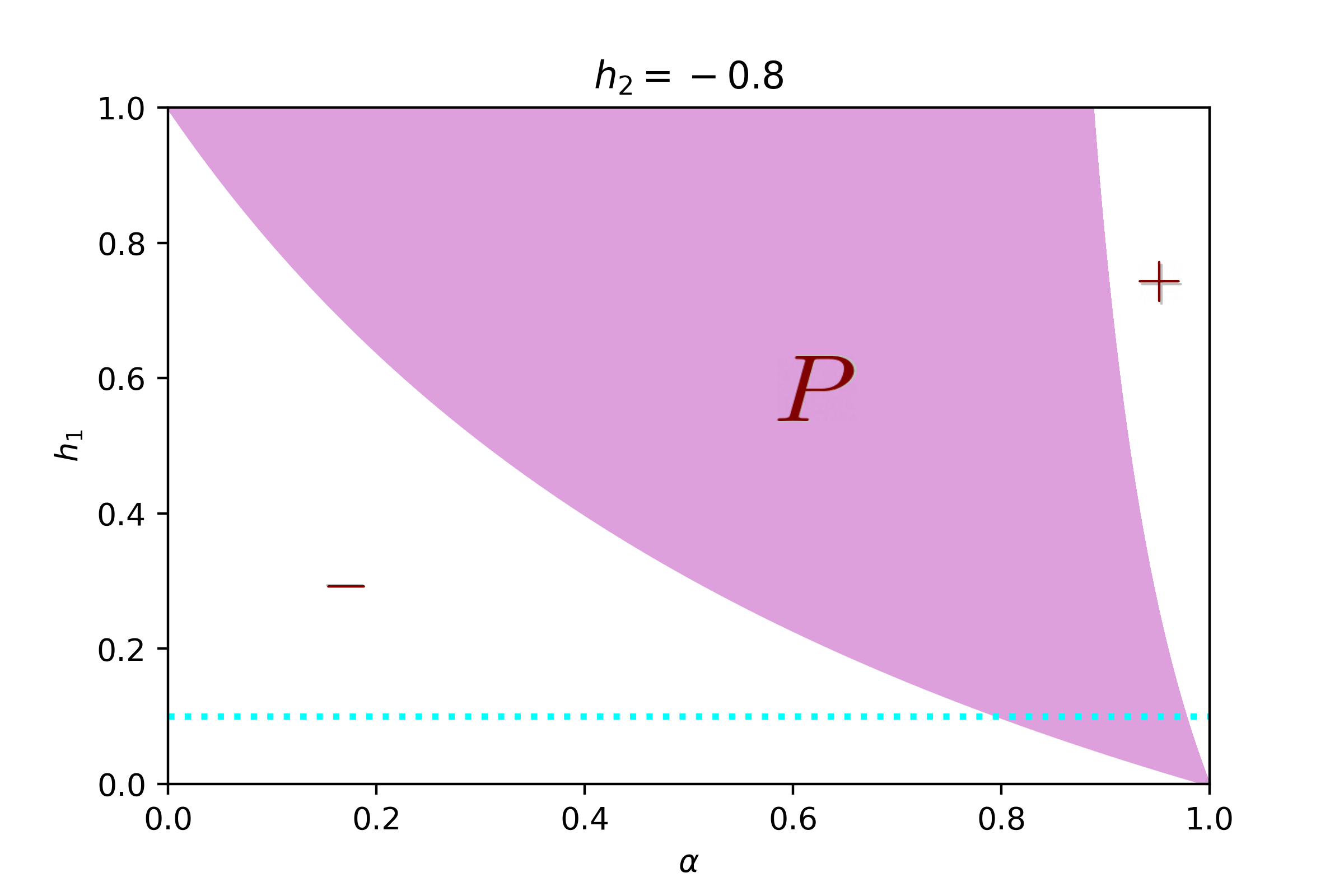}} 
\subfloat(d){\includegraphics[width = 2.7in]{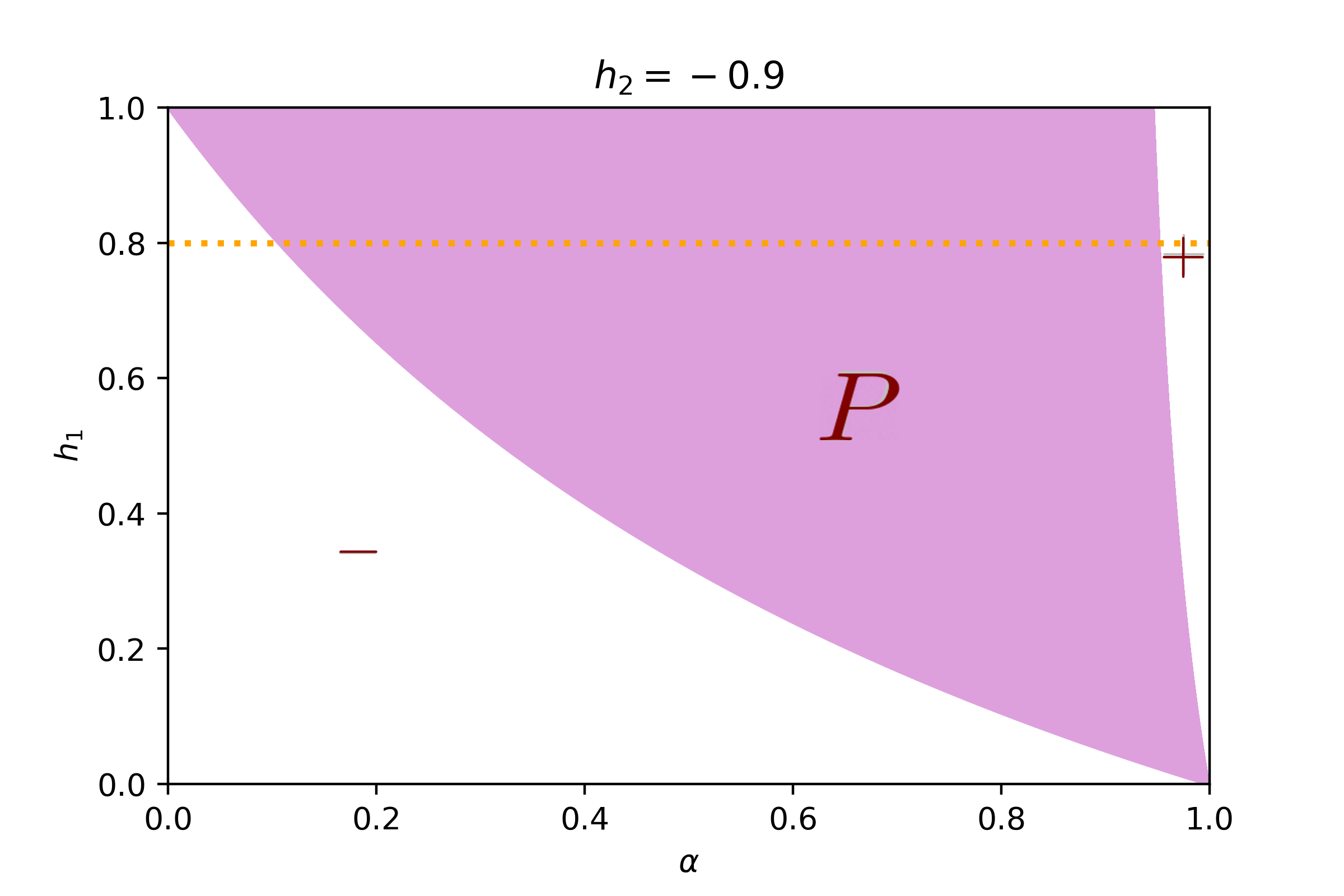}}\\
\caption{\textbf{Phase diagrams for the complete network.} The polarized area is colored in light purple and indicated with $P$, in white are $+$ and $-$ consensus. Figure (a) shows the mean-field phases in the $\alpha h$ plane, for equal and opposite preferences' intensities $h_1=-h_2=h$. The dark purple dot represents the regime $\alpha =\frac{1}{2},h\rightarrow0$ in which \cite{borile2013effect} have investigated finite size effects. Figures (b),(c),(d) report the phases in the $\alpha h_1$ plane, once fixed $h_2$ respectively to $-0.6,-0.8,-0.9$. Each of the colored horizontal lines present in some of the plots represents the choice of the biases $h_1,h_2$ as in figure \ref{fig:fc_bifurcation} (the colors correspond). They are reported in order to show how the lines intersecate the different phases.}
\label{fig:fc phase diagrams}
\end{figure}

First we study the model, for simplicity, on the complete network. This setting was already investigated in \cite{masuda2011can}, however, we complete the analysis by calculating the polarization measure at the stationary state for any choice of the parameters.\\
Let $\alpha=\frac{N_1}{N}$ be the fraction of nodes in the first class characterized by bias $h_1$. We define
\begin{equation}
    \rho_1 = \frac{\sum_{i=1}^{N_1}\frac{1+\sigma_i}{2}}{N} \in [0,\alpha]\;\;\;\;\;\;\;\;\;\;\;  \rho_2 = \frac{\sum_{i=N_1+1}^{N}\frac{1+\sigma_i}{2}}{N}\in[0,1-\alpha]
\end{equation}
as the ratios between the number of first class spins (respectively second) in current up $+1$ state and the total number of spins in the system. The system of coupled ordinary differential equations which describes the evolution of such dynamical variables can be written generally in terms of the global rates $R_{\pm1/2}$
\begin{equation}
    \begin{cases}
    \dot{\rho_1} = R_{+1}(\rho_1,\rho_2) - R_{-1}(\rho_1,\rho_2)   \\
    \dot{\rho_2} = R_{+2}(\rho_1,\rho_2) - R_{-2}(\rho_1,\rho_2)
    \end{cases}
    \label{eq:rate}
\end{equation}
For example, the global rate $R_{+1}$ represents the probability per unit time that, when the system is currently in state $\rho_1,\rho_2$, a spin of the first class undergoes the transition $-1\rightarrow+1$, increasing 
the density of up spin of the first class of $1/N$, $\rho_1\rightarrow \rho_1+\frac{1}{N}$. Considering a time unit corresponding to $N$ steps, i.e. $\delta t= N^{-1}$, the transition rates for a fully connected network are
\begin{equation}
    \begin{cases} 
    R_{+1}(\rho_1,\rho_2) = (\alpha-\rho_1)\frac{1+h_1}{2}(\rho_1+\rho_2) \\
    R_{-1}(\rho_1,\rho_2) =  \rho_1\frac{1-h_1}{2}(1-\rho_1 -\rho_2)\\\\
    R_{+2}(\rho_1,\rho_2) =  (1-\alpha-\rho_2)\frac{1+h_2}{2}(\rho_1+\rho_2)\\
    R_{-2}(\rho_1,\rho_2) =  \rho_2\frac{1-h_2}{2}(1-\rho_1-\rho_2)
    \end{cases}
\end{equation}
For example, in the first rate $R_{+1}$ the first term $\alpha-\rho_1$ is the probability of chosing uniformly randomly a spin of the first class currently in down state, while $\rho_1+\rho_2$ is the probalbility of choosing a neighbour in state $+1$ in the complete network, and eventually $\frac{1+h_1}{2}$ is the probability of the transition, according to the model dynamics. Thus we have the following mean-field equations
\begin{equation}
\begin{cases}
    \dot{\rho_1} = \frac{1}{2}\bigg[   (\alpha-\rho_1)(1+h_1)(\rho_1+\rho_2) - \rho_1(1-h_1)(1-(\rho_1+\rho_2))\bigg]  \\
    \dot{\rho_2} = \frac{1}{2}\bigg[ (1-\alpha-\rho_2)(1+h_2)(\rho_1+\rho_2) - \rho_2(1-h_2)(1-(\rho_1+\rho_2)) \bigg] 
\end{cases}
\label{fc mf rho12}
\end{equation}
For the complete network in the $N\rightarrow\infty$ limit, the mean-field equations represent exactly the evolution of the system and they
can be applied, as an approximation, 
to other networks. Not considering structural or dynamical correlations, we expect them to be still accurate on a sufficiently dense network without 
specific structural features \cite{peralta2021effect,porter2014dynamical}, such as an Erd\H os-R\'enyi random graph with probability of linkage of $O(1)$. \\
Localizing the fixed points $(\rho_1^*,\rho_2^*)$ of the system (\ref{fc mf rho12}) and characterizing their stability by the analysis of the corresponding Jacobian matrices reported in the appendix, one finds \cite{masuda2011can} that 
\begin{itemize}
    \item The positive $(1,1)$ (all up spins) and negative $(0,0)$ (all down spins) consensus points are always fixed points, for any combination of the parameters $\alpha,h_1,h_2$.
    \item When the positive (or negative) consensus is stable, it is the only stable fixed point.
    \item When both the consensus fixed points are not stable, another fixed point with $\rho_1^*,\rho_2^*\in(0,1)$ appears. Such fixed point, when it exists, is always stable.
\end{itemize}
Defining the total density of up spin $\rho= \rho_1 + \rho_2$ and $\Delta = \frac{\rho_1}{\alpha} - \frac{\rho_2}{1-\alpha} $, the polarization can be expressed\footnote{If we consider the group magnetizations $m_1=\sum\limits_{i=1}^{N_1}\sigma_i$, $m_2=\sum\limits_{i=N_1+1}^{N}\sigma_i$ it is easy to reconduct the expression before to the more conventional expression of the polarization $P=\frac{|m_1-m_2|}{2}\in[0,1]$} as $P=|\Delta|$. As a first contribution of this paper, we calculate that at the \textit{impasse} or \textit{polarized} state the average density of up spins and the polarization respectively read
\begin{align}
    &\rho^* = \frac{1}{2} \bigg(\frac{h_2-h_1}{h_1h_2}\alpha - \frac{1-h_1}{h_1}\bigg) \label{pol_rho} \\
    &\Delta^* =   \frac{1}{h_1-h_2}\bigg( 1 + \frac{\alpha^2h_1^2+(1-\alpha)^2h_2^2-h_1^2h_2^2}{2\alpha(1-\alpha)h_1h_2}\bigg)\label{pol_delta}
\end{align}
Moreover, by analyzing the Jacobian one can localize the critical value of the parameters at which the transitions from negative consensus to polarization and from polarization to positive consensus occur: taking $h_1,h_2$ fixed and letting $\alpha$ vary, we have that the critical points of the transitions above are respectively at
\begin{align*}
        &\alpha^-_c = (1-h_1)\frac{h_2}{h_2-h_1} \\
        &\alpha^+_c = (1+h_1)\frac{h_2}{h_2-h_1}  
\end{align*}
The bifurcation diagrams, taking $\rho$ and $P$ as order parameter and varying the composition $\alpha$ for various fixed $h_1,h_2$, are shown in Figure \ref{fig:fc_bifurcation}: the bifurcation is of transcritical type and the transitions are indeed continous. The presented numerical simulations confirm that the analytical solutions work well for systems defined on relatively small, $N=1000$ complete graphs as well.
\\ 
The length of the interval associated to the polarized state is $\alpha^+_c - \alpha^-_c =2\frac{h_1h_2}{h_2-h_1}$ which reduces to $\alpha^+_c - \alpha^-_c = h$ for $h_1=-h_2 = h$. The phase diagram in this case (in the $\alpha, h$ plane) is shown in Figure \ref{fig:fc phase diagrams}a, while in the remaining plots of the figure $h_2$ is fixed to different values and the phases in the plane $\alpha, h_1$ are shown. To link the bifurcation and the phase diagrams, the horizontal lines corresponding to the choices of the biases in figure \ref{fig:fc_bifurcation} are reported on the latters.\\
Defining the critical mass of a population \cite{centola2018experimental} as the minimum fraction of individuals of that population necessary to escape from consensus at the unpreferred opinion, we have that the critical masses of respectively the first and second populations are $\alpha^{-}_c$ and $1-\alpha_c^{+}$. For very low biases, the populations over the critical masses rapidly overturn the outcome of the system, switching the direction of consensus. In this model, the critical masses depend only on the biases and lay in the whole range $(0,1)$.

\section{VMP on modular networks}
\label{chapter SBM}
\begin{figure}
    \centering
    \includegraphics[width=0.7\textwidth]{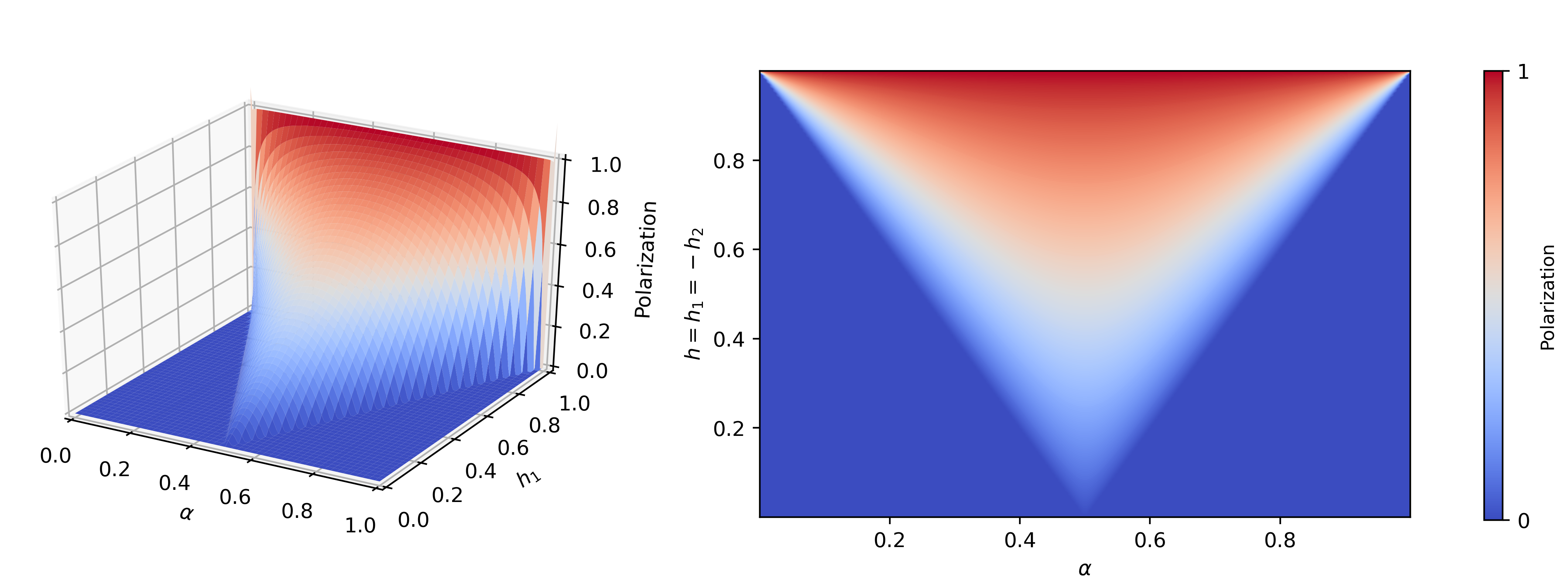}\\
    \includegraphics[width=0.7\textwidth]{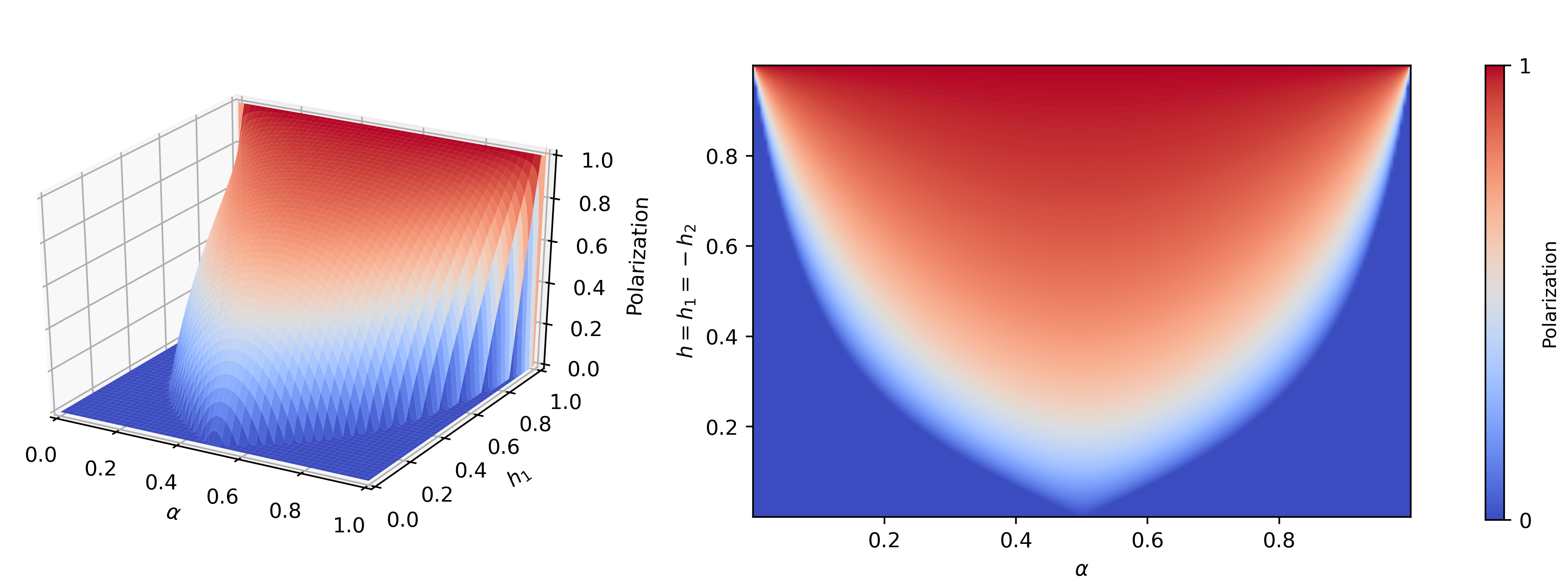}
    \caption{\textbf{VMP on modular networks with equally strong opposite preferences, $\alpha, h$ plane.} The mean field predictions are shown, i.e. the stable fixed point of the system (\ref{SBM mf system}), for two choices of connectivities of the modular network. \textit{Upper plots}: case $p_{11}=p_{22}=p_{12}=p_{21}=1$, corresponding to the fully connected topology (subplot (a) in figure \ref{fig:fc phase diagrams}). \textit{Lower plots}: case $p_{11}=p_{22}=1$, $p_{12}=p_{21}=0.3$, i.e. still symmetric probabilities but segregated communities. The polarization increases and the polarization area widens, nevertheless the general qualitative behaviour is preserved.}
    \label{fig:SBMalpha}
\end{figure}

\begin{figure}
    \centering
    \includegraphics[width=\textwidth]{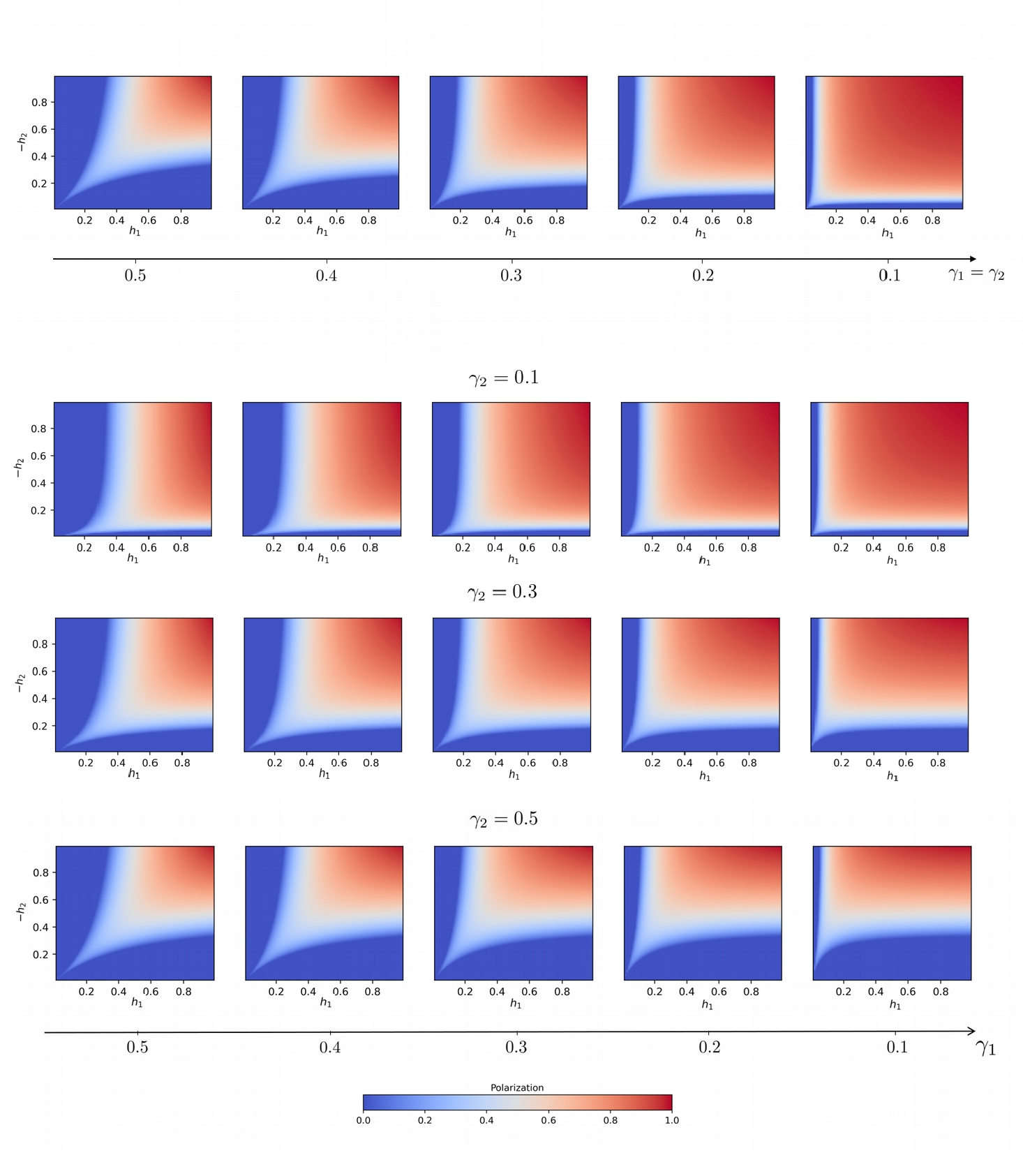}
    \caption{\textbf{VMP on modular networks.} Mean-field predictions for the polarization (eq. \ref{mf_echochambers}) in the $h_1,h_2$ plane for two equally open-minded communities ($\gamma_1=\gamma_2$, upper line), and for three values of $\gamma_2$, varying the open-mindedness of the first population $\gamma_1$. Decreasing the open-mindedness, the consensus areas shrink and the system becomes more and more polarized.}
    \label{fig:echo_gamma}
\end{figure}

\begin{figure}
    \textit{\subfloat(a)}{\includegraphics[width = 3in]{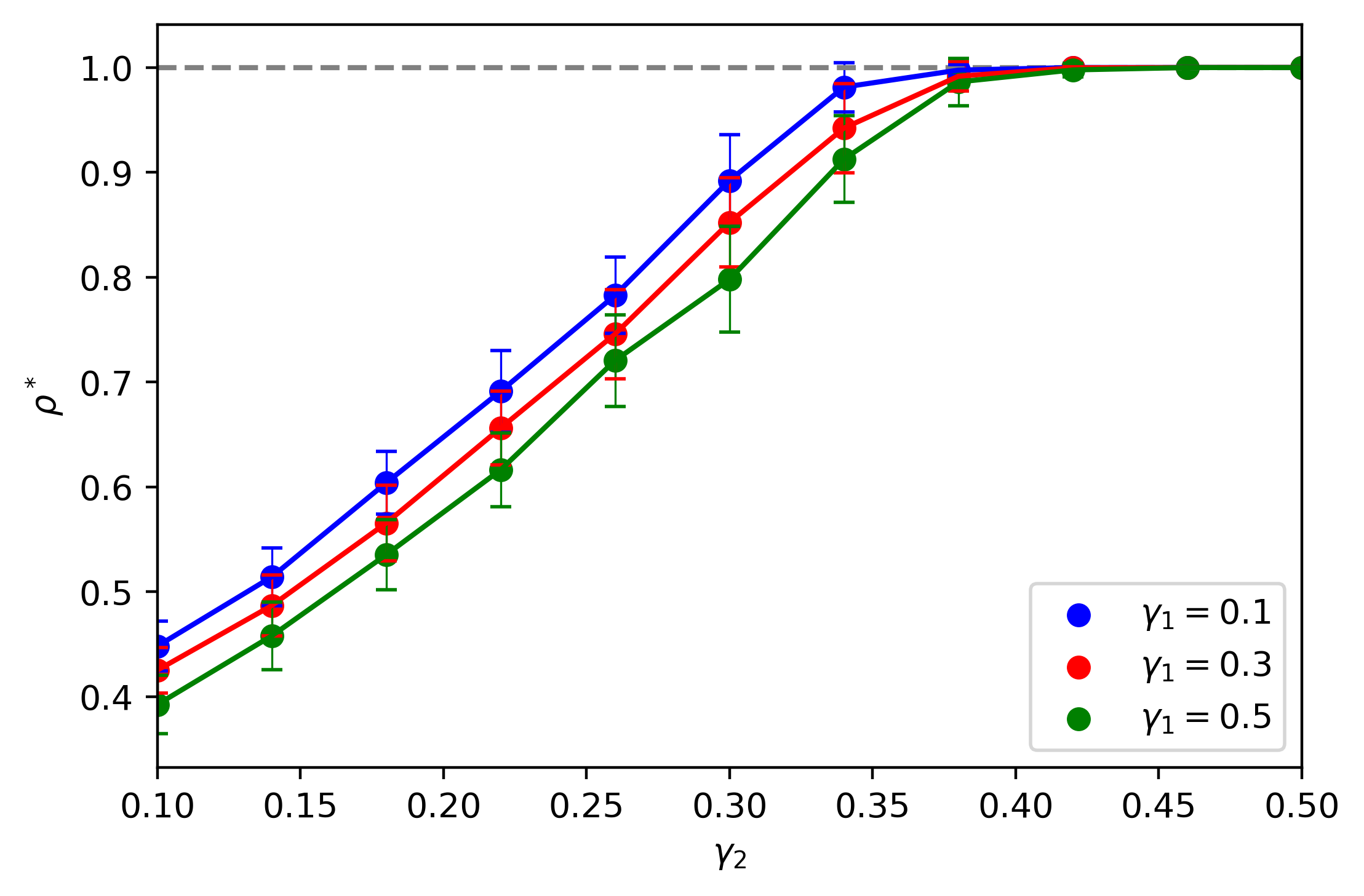}}
    \textit{\subfloat(b)}{\includegraphics[scale = 0.12]{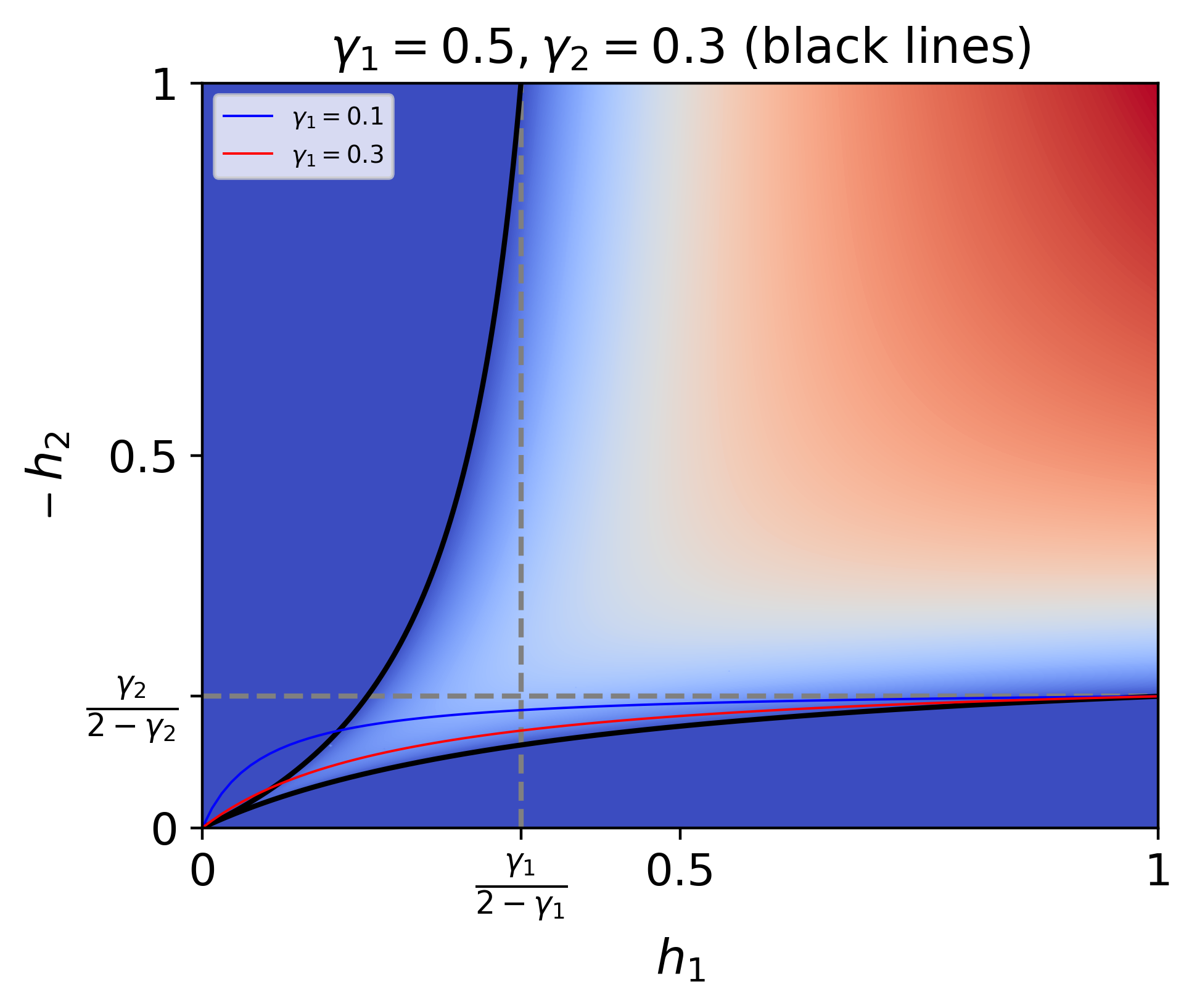}}\\
    \textit{\subfloat(c)}{\includegraphics[width = 3in]{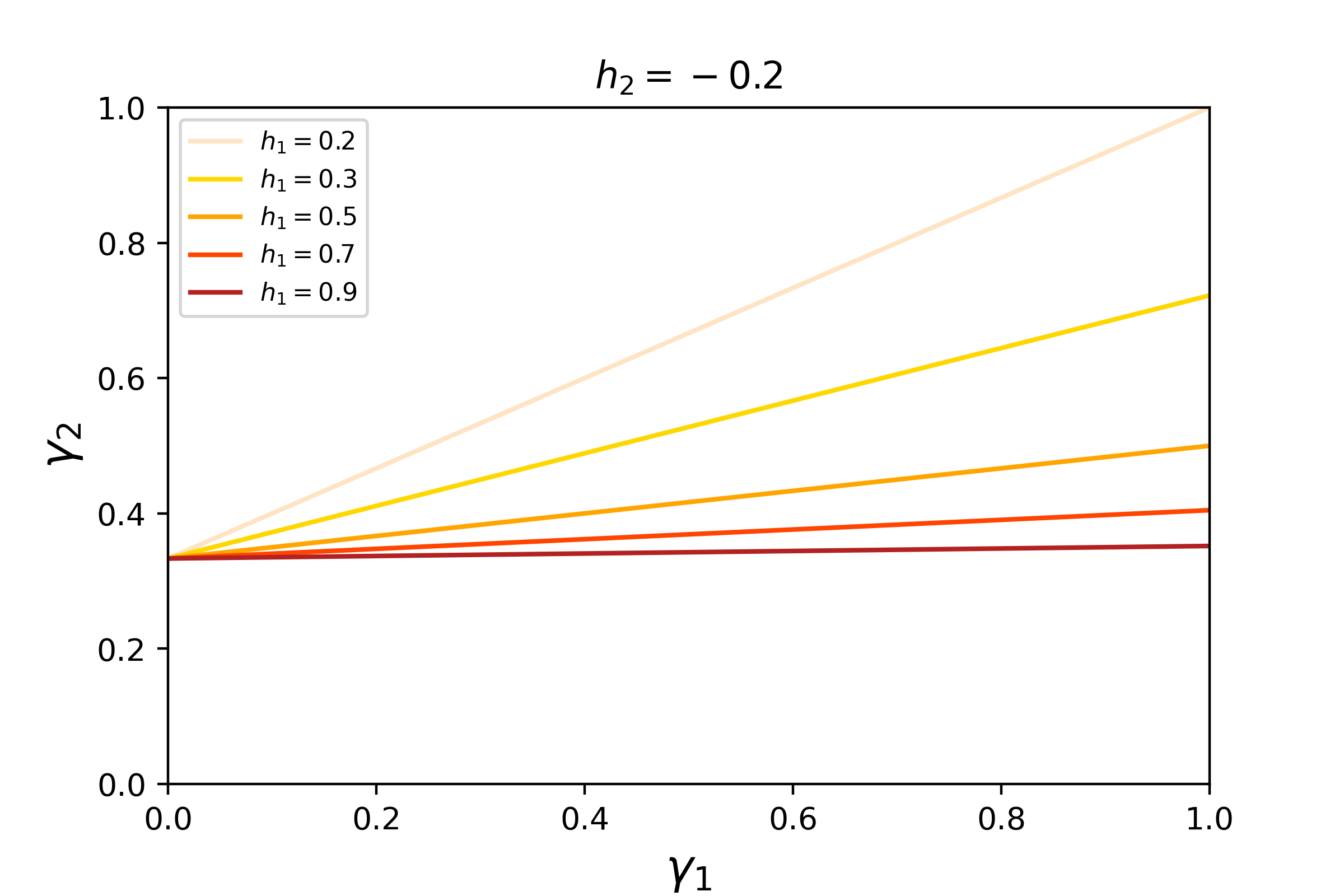}} 
    \textit{\subfloat(d)}{\includegraphics[width = 3in]{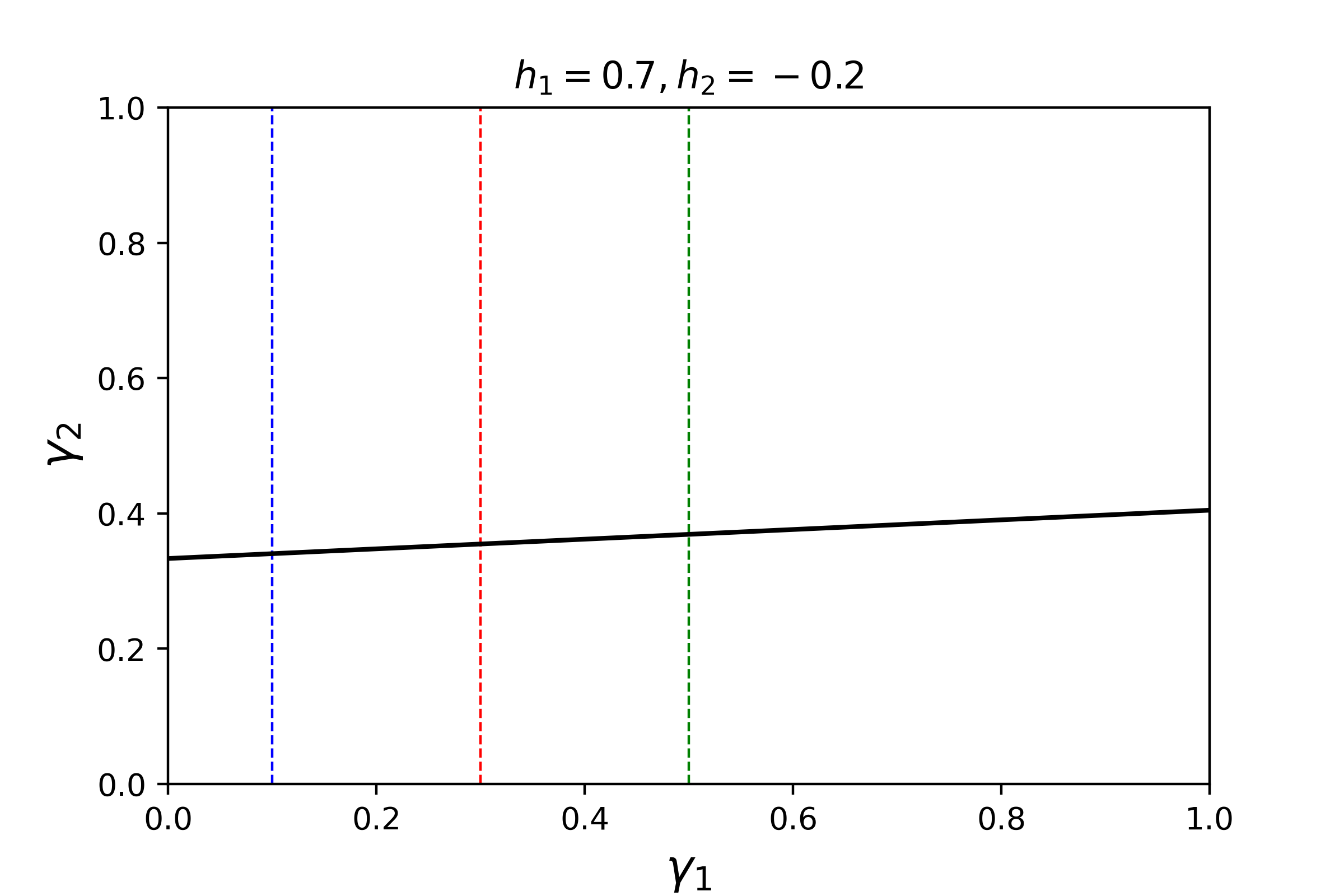}}\\
    \caption{\textbf{Simulations and analytical results for the VMP on a modular graph.}\\ \textit{(a)} The average density of up spin at the stationary state is calculated over 30 runs of the model, with $N=1000$ agents and preferences $h_1=0.7,h_2=-0.2$, varying the open-mindedess of the second population and for 3 values of the open-mindedness of the first population, corresponding to different colors. We can see that all the three lines approach the positive consensus at approximately the same $\gamma_2$. \\
    \textit{(b)} Critical lines for the positive and negative consensus points in the $h_1h_2$ plane, for $\gamma_1=0.5,\gamma_2=0.3$ (black). On the background, the polarization values calculated numerically as in figure \ref{fig:echo_gamma} are reported, to show the consistency of the results of the linear stability analysis. The red and blue lines are the positive consensus critical lines for other values of $\gamma_1$, while $\gamma_2=0.3$ constantly. The figure is the analytical correspondant of $\gamma_1 = 0.1,0.3,0.5$, $\gamma_2=0.3$ plots in figure \ref{fig:echo_gamma}. \\
    \textit{(c)} Critical lines in the $\gamma_1\gamma_2$ plane for $h_2=-0.2$ and different $h_1$. The figure shows that the higher $h_1$, the more the critical line tends to an horizontal line. \\
    \textit{(d)}  Once fixed the biases to $h_1=0.7,h_2=-0.2$, in the $\gamma_1\gamma_2$ plane are reported the analytical critical line (\ref{lsa critical line gamma}) and the values of $\gamma_1$ (vertical lines) corresponding to the lines in figure (a): we see that the intersections with the critical line (i.e. the points at which the positive consensus becomes stable) occur almost at the same $\gamma_2$, for all the three values of $\gamma_1$.    }
    \label{fig:linear_stab_analysis}
\end{figure}

To study the effect of the topology reflecting the biased communities of the bipopulated VMP, we analyze the model on a network with two modules of sizes $N_1$ and $N_2$, generated by a Stochastic Block Model (SBM) \cite{lee2019review}. The SBM is defined by the intra-modular ($p_{11},p_{22}$) and intermodular ($p_{12},p_{21}$) connectivities, i.e., the probabilities describing the corresponding linkings between the agents (for undirected networks $p_{12}=p_{21})$). We assume that the network results from homophilic interactions (epistemic bubbles) such that all agents within module $1$ ($2$) have bias $h_1$ ($h_2$).
\\
We start from the mean-field equations (\ref{eq:rate}) where global transition rates are now functions of the connectivities of the block model
\begin{equation}
    \begin{cases} 
    R_{+1}(\rho_1,\rho_2) =\frac{1}{\alpha p_{11} + (1-\alpha)p_{12}} (\alpha-\rho_1)\frac{1+h_1}{2}(p_{11}\rho_1+p_{12}\rho_2) \\
    R_{-1}(\rho_1,\rho_2) = \frac{1}{\alpha p_{11} + (1-\alpha)p_{12}} \rho_1\frac{1-h_1}{2}[p_{11}(\alpha-\rho_1)+p_{12}(1-\alpha -\rho_2)]\\\\
    R_{+2}(\rho_1,\rho_2) = \frac{1}{\alpha p_{12} + (1-\alpha)p_{22}} (1-\alpha-\rho_2)\frac{1+h_2}{2}(p_{12}\rho_1+p_{22}\rho_2)\\
    R_{-2}(\rho_1,\rho_2) = \frac{1}{\alpha p_{12} + (1-\alpha)p_{22}} \rho_2\frac{1-h_2}{2}[p_{12}(\alpha-\rho_1)+p_{22}(1-\alpha -\rho_2)]
    \end{cases}
    \label{SBM transition rates}
\end{equation}
For example, once a spin in down state of class 1 is selected, the probability that we find one of its neighbours in up state is 
$\frac{p_{11}\rho_1}{\alpha p_{11} + (1-\alpha)p_{12}}+\frac{p_{12}\rho_2}{\alpha p_{11} + (1-\alpha)p_{12}}$ (for more details, see the Appendix \ref{app transition rates}).\\
We end up with the mean-field evolution equations for the densities $\rho_1,\rho_2$
\begin{equation}
    \begin{cases}
    \dot{\rho_1} = C_1 \bigg[ (\alpha-\rho_1)(1+h_1)(p_{11}\rho_1+p_{12}\rho_2) - \rho_1(1-h_1)[p_{11}(\alpha-\rho_1)+p_{12}(1-\alpha -\rho_2)] \bigg]  \\
    \dot{\rho_2} = C_2 \bigg[ (1-\alpha-\rho_2)(1+h_2)(p_{12}\rho_1+p_{22}\rho_2) - \rho_2(1-h_2)[p_{12}(\alpha-\rho_1)+p_{22}(1-\alpha -\rho_2)] \bigg] 
    \end{cases}
\label{SBM mf system}
\end{equation}
where $C_1 = \frac{1}{2[\alpha p_{11} + (1-\alpha)p_{12}]}$, $C_2 = \frac{1}{2[\alpha p_{12} + (1-\alpha)p_{22}]}$.\\
It is easy to see that the consensus states are still fixed points. The numerical study of the system shows that the qualitative behaviour of the fully connected case is preserved, i.e., the different phases are separated by transcritical bifurcations, but the polarization area generally widens and the range of the polarized phase increases. Figure \ref{fig:SBMalpha}
compares the fully connected case and the topology characterized by two cliques ($p_{11}=p_{22}=1$) and intercommunity connectivities $p_{12}=p_{21}=0.3$, with equally strong opposite preferences $h_1=-h_2=h$. In general, the symmetric community structure decreases the values of the critical masses, with respect to the fully connected topology.
\newpage
By considering the evolution of the normalized densities $\rho'_1 = \frac{\rho_1}{\alpha} \in [0,1]$, $\rho'_2 = \frac{\rho_2}{1-\alpha} \in [0,1]$ and defining the topological parameters
\begin{equation}
    \gamma_1 = \frac{p_{12}(1-\alpha)}{p_{11}\alpha +p_{12}(1-\alpha)} \;\;\;\;\;\;\;\;\; \gamma_2 = \frac{p_{21}\alpha}{p_{22}(1-\alpha) +p_{21}\alpha}
\end{equation}
we have a consistent reduction in the number of parameters in the mean-field system on the SBM (\ref{SBM mf system}), which reads
\begin{equation}
\begin{cases}
    \dot{\rho'_1} = \frac{1}{2}\bigg[   (1-\rho'_1)(1+h_1)[(1-\gamma_1)\rho'_1+\gamma_1\rho'_2] - \rho'_1(1-h_1)[(1-\gamma_1)(1-\rho'_1)+\gamma_1(1-\rho'_2)]\bigg]  \\
    \dot{\rho'_2} = \frac{1}{2}\bigg[ (1-\rho'_2)(1+h_2)[\gamma_2\rho'_1+(1-\gamma_2)\rho'_2) - \rho'_2(1-h_2)[\gamma_2(1-\rho'_1)+(1-\gamma_2)(1-\rho'_2)] \bigg] 
\end{cases}
\label{mf_echochambers}
\end{equation}
The interpretation of $\gamma_1,\gamma_2$ is straightforward in terms of the average internal and external degrees $z_{11},z_{12},z_{21},z_{22}$  \footnote{The internal degrees $z_{11},z_{22}$ are the numbers of connections that an agent of respectively the first and second community on average has within his community, while the external degrees $z_{12},z_{21}$ are the average number of connections towards the other community}
\begin{equation}
    \gamma_1 = \frac{z_{12}}{z_{11}+z_{12}} \;\;\;\;\;\;\;\;\; \gamma_2 = \frac{z_{21}}{z_{22}+z_{21}}
\end{equation}
Being the average fractions of external connections over the total number of connections of the agents in class 1 and 2, $\gamma_1,\gamma_2$ can be intended as the average \textit{open-mindedness} of the individuals of respectively the first and second community. Since in an epistemic bubble the agents overrepresent (i.e., are more linked to) their belonging community, we expect $\gamma_1 \in (0,1-\alpha)$ and $\gamma_2 \in (0,\alpha)$. The more far $\gamma_1,\gamma_2$ are from these upper extremes, the more the individuals of the corresponding population have unbalanced sources of information, i.e. are trapped in the bubble. If the individuals have on average more connections within their belonging community rather than towards the other, then $\gamma_1,\gamma_2\in(0,0.5)$.
Figure \ref{fig:echo_gamma} shows the stationary polarization\footnote{$P^*=\Delta^*$, since it is easy to check that at the fixed point $\Delta^*\geq0$ for $h_1\geq0,\;h_2\leq0$.} $\Delta^* = \rho'^*_1-\rho'^*_2$ in the $h_1h_2$ plane, for different choices of the open-mindedness parameters $\gamma_1,\gamma_2$, by numerically calculating the fixed points of the mean-field system (\ref{mf_echochambers}) and determining their stability.\\
\\
\\
From the linear stability analysis of the consensus fixed points of the system (\ref{mf_echochambers}), reported in Appendix \ref{Modular lsa}, we obtain the condition for the stability of the positive consensus 
\begin{equation}
    \gamma_1h_2(1-h_1)+\gamma_2h_1(1-h_2) + 2h_1h_2\geq0
\end{equation}
and for the negative one
\begin{equation}
     -\gamma_1h_2(1+h_1)-\gamma_2h_1(1+h_2)+2h_1h_2\geq0
\end{equation}
First we fix $\gamma_1,\gamma_2$ and determine the critical lines in the $h_1,h_2$ plane. The line separating the space when the positive consensus is stable (below the line) and the one for which it is unstable reads
\begin{equation}
    -h_2^c(h_1) = \frac{\gamma_2h_1}{\gamma_1(1-h_1)+h_1(2-\gamma_2)}
\end{equation}
and it is bounded superiorly by $-h_2^{c+} = \frac{\gamma_2}{2-\gamma_2}$, which is reached at $h_1=1$ and does not depend on $\gamma_1$. \\
In the same way, the critical line related to the negative consensus fixed point reads
\begin{equation}
    h_1^c(h_2) = \frac{-\gamma_2h_2}{\gamma_2(1+h_2)-h_2(2-\gamma_1)}
\end{equation}
and it is bounded by $h_1^{c-} = \frac{\gamma_1}{2-\gamma_1}$, independent of $\gamma_2$. The critical lines are drawn for a choice of the open-mindedness parameters in figure \ref{fig:linear_stab_analysis}b.\\
In the $\gamma_1,\gamma_2$ plane, the critical line  for the positive consensus
\begin{equation}
    \gamma_2^c(\gamma_1) = \frac{-h_2(1-h_1)}{h_1(1-h_2)} \gamma_1 -\frac{2h_2}{(1-h_2)}
\label{lsa critical line gamma}
\end{equation}
is a straight line in the plane, whose coefficient tends to zero (the line flattens) for high $h_1$ and low $-h_2$. When this is the case, the critical points for different topologies of the first population, i.e. different $\gamma_1$, happen to be at approximately the same value of $\gamma_2$, as shown in figures \ref{fig:linear_stab_analysis}c and \ref{fig:linear_stab_analysis}d. Naturally, symmetric considerations hold for the negative consensus.\\
\\
The results of linear stability analysis allow us to conclude that if one population is very committed and the other population's bias is low, e.g. when $h_1>>-h_2$, wether or not such population manages to impose its preferred opinion depends largely on the open-mindedness of the other population $\gamma_2$,  while the dependency on the topological structure of the committed population $\gamma_1$ is negligible.\\
\\
We test this claim by numerically simulating the VMP on modular networks for a finite system of $N=1000$ agents, fixing $h_1=0.7,h_2=-0.2$ and varying the topology through the open-mindedness parameters $\gamma_2$ and $\gamma_1$. The results, shown in Figure \ref{fig:linear_stab_analysis}a, support the mean-field predictions by showing that the critical value of $\gamma_2$ separating the polarization and positive consensus phases is approximately the same for all the three lines in the $\gamma_2,\rho^*$ plane, corresponding to the three chosen values of $\gamma_1$.

\section{Pair Approximation}\label{PA_chapter}
\begin{figure}
\subfloat{\includegraphics[width = 0.5\textwidth]{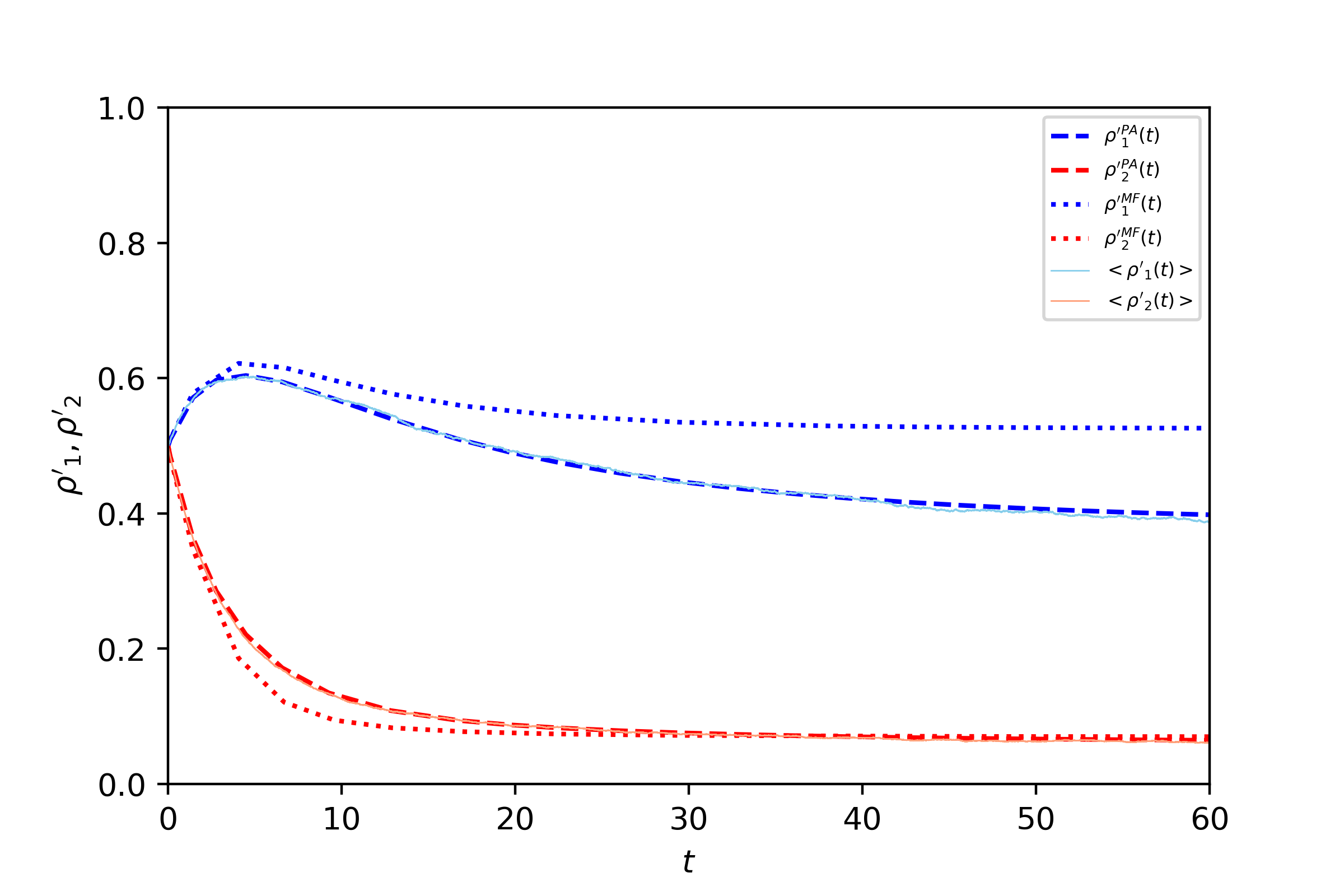}} 
\subfloat{\includegraphics[width = 0.5\textwidth]{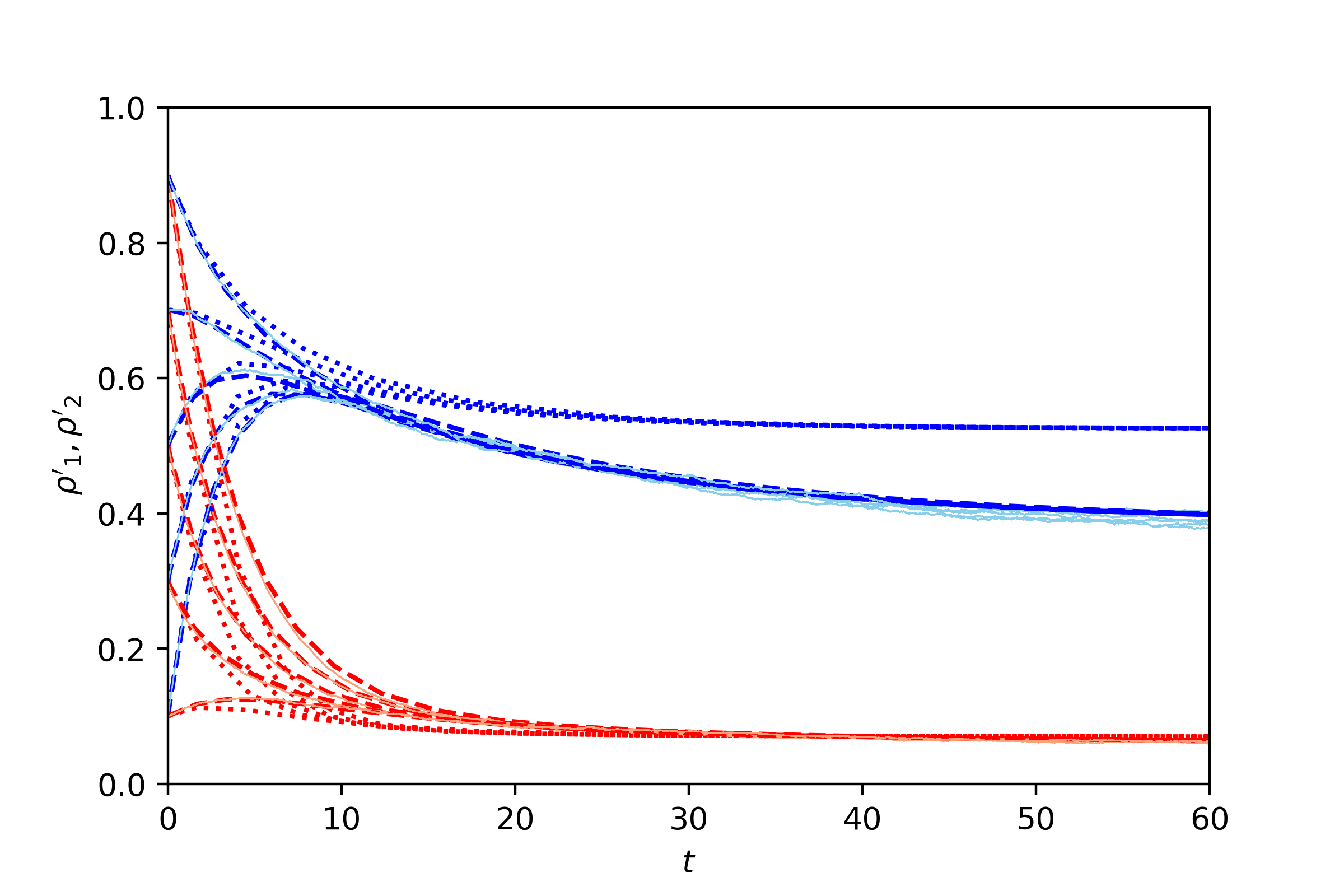}}\\
\caption{\textbf{Pair and Mean-Field approximations.} \textit{Left plot}: Dynamics of the MF (dotted line) and PA (dashed lines) approximations compared with numerical simulations (thin solid lines) averaged over $30$ independent runs. The graph is a sparse modular network of $N=3000$ nodes with two $z-$regular communities $z_{11}=4,z_{12}=2,z_{21}=1,z_{22}=3$, thus $\alpha = \frac{2}{3}$ and $\gamma_1=0.4 ,\gamma_2=0.25$. The two communities' preferences are respectively $h_1=0.3,h_2=-0.5$. The initial opinions are chosen uniformly random ($\rho_1'(0) = \rho_2'(0) = \frac{1}{2}$). \textit{Right plot}: Same setting repeated for various initial conditions $\rho'_1(0),\rho'_2(0)$. The results show that even for low connectivities the stable fixed point is unique.}
\label{fig:PA and MF}
\end{figure}

The aim of this section is to investigate how a more refined approximation better reproduces the model's behaviour on complex networks. We implement the so called Pair Approximation \cite{gleeson2011high,gleeson2013binary}, which takes into consideration dynamical correlations at a pairwise level. \\
For the sake of simplicity, the Pair Approximation is applied on an undirected regular\footnote{regular in the sense that each node of the class has the same number of internal $z_{ii}$ and external $z_{ij}$ connections.} modular graphs with two communities, nevertheless the treatment can be easily extended to a directed modular network with heterogeneous degrees \cite{pugliese2009heterogeneous}. Generalizing the approach followed by \cite{peralta2021effect} to class-dependent infection/recovery rates, reported in the appendix \ref{app_PA}, we consider the set of probabilities of the kind $P^{ij\pm}_{z_{ij},m_{ij}}$, i.e. the probability that a node in the population $i$ currently in state $\pm1$ has $z_{ij}$  degree\footnote{trivial in the $z$-regular case, explicited only for the generalization} and $m_{ij}$ neighbours of the population $j$ currently in state $+1$.\\
The mean-field approximation consists in taking those probabilities highly peaked at the values of $\frac{m_{ij}}{z_{ij}}$ coinciding to the overall normalized densities $\rho'_1,\rho'_2$, so having the shape of delta functions explicited in (\ref{MF_probabilities_start}-\ref{MF_probabilities_end}). The Pair Approximation, instead, considers pairwise dynamical correlations by taking those probabilities as binomial distributions with single event probabilitiy corresponding to the ratio between the number of active links departing from the node's type, intended as its class and state, and the number of connections from that node's type. The argument is explained in the appendix (\ref{eq:PA_proba_start}-\ref{PA_proba_end},\ref{PA_single_proba_start}-\ref{PA_single_proba_end}).\\
The latter ratios are also dynamical variables of the system derived from the pair approximation (\ref{PA_system}), which consists indeed of six coupled differential equations. The increase in complexity is justified by a gain in accuracy, consistent for low average degrees, i.e. for sparse graphs. In figure \ref{fig:PA and MF} we compare the mean-field and Pair approximation effectiveness in reproducing the dynamics of the system, on an extremely sparse modular network. We see that the gain in accuracy is consistent, and the PA is able to predict the dynamics almost perfectly. On the right plot of the figure, we test multiple initial conditions in order to check that, as in the mean-field treatment, the stable fixed point is unique. 

\section{Application to the network of blogs}\label{blognetwork_chapter}
\begin{figure}
\centering
\subfloat{\includegraphics[width = 0.8\textwidth]{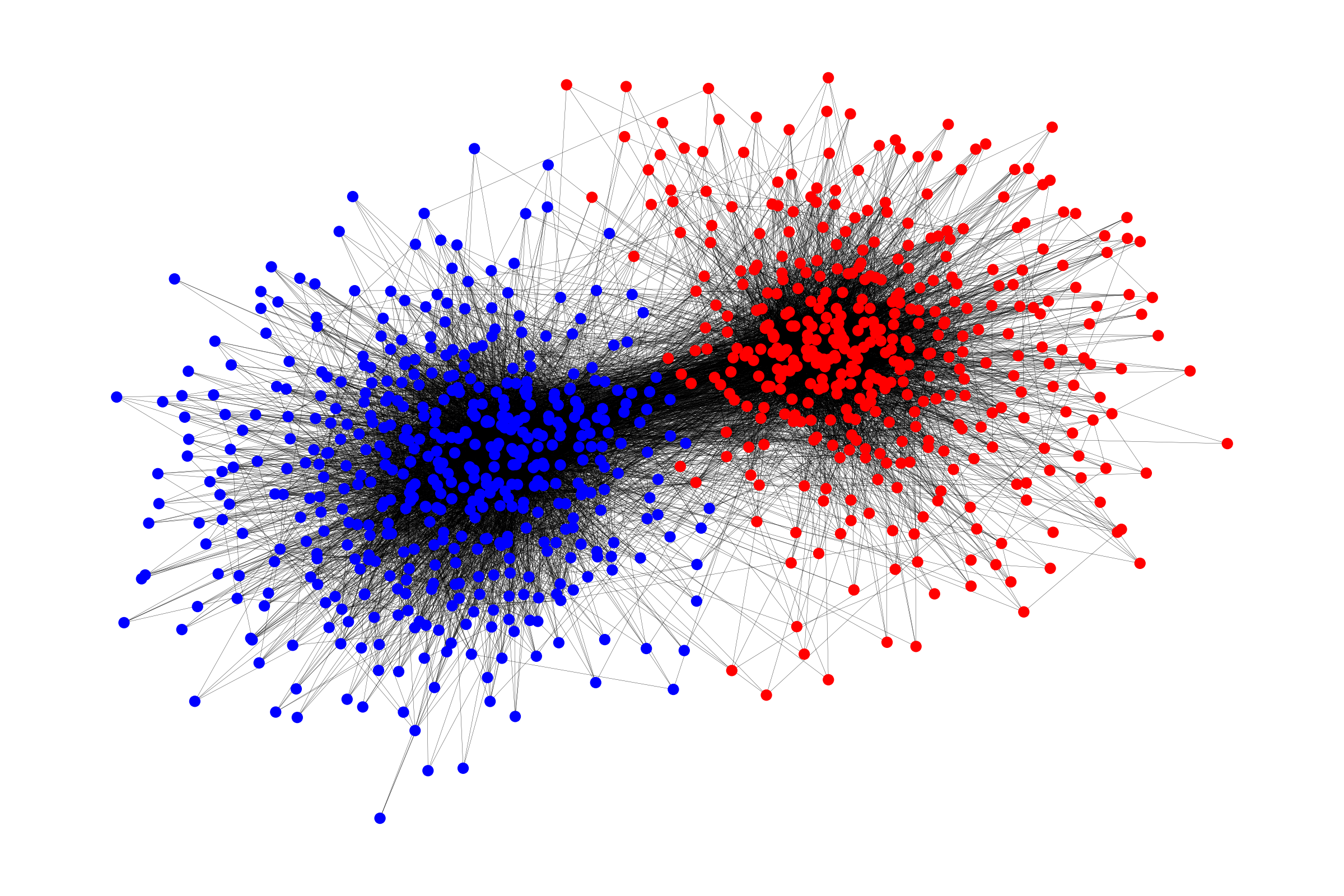}}\\ 
\subfloat{\includegraphics[width = 0.8\textwidth]{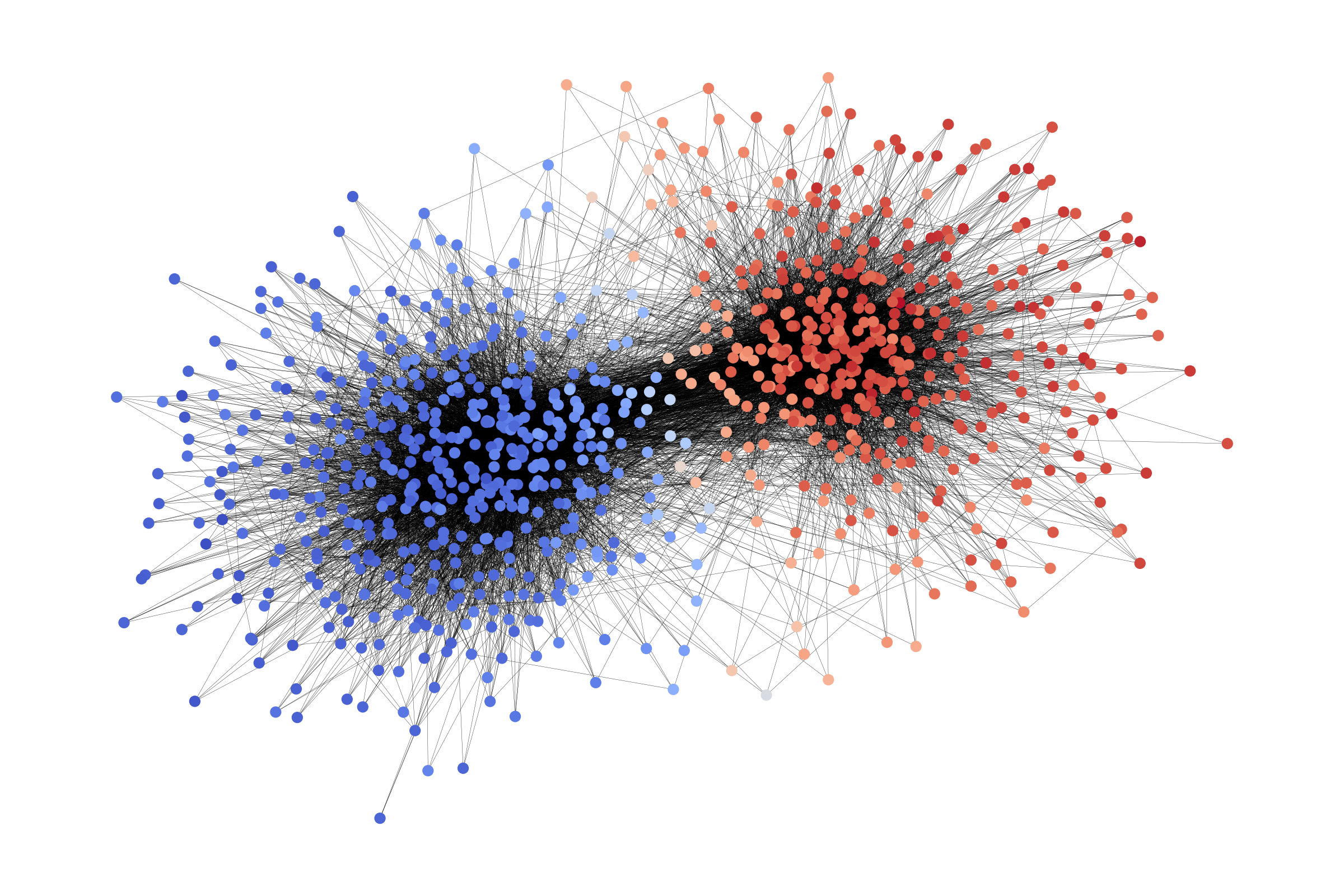}}
\caption{\textbf{VMP on the 2004 Blogosphere.} \textit{Upper plot}: 2004 Political Blogosphere after community detection: nodes' colors reflect political partisanships. \textit{Lower plot}: each node is colored according to the average state $\bar{\rho_i}$ at time $t=50$, for a bipopulated VMP where the populations are the ones computed by community detection and the preferences are set to $h_r = -h_b = 0.2$. The colorscale goes from blue, corresponding to $\bar{\rho_i}=0$, to red ($\bar{\rho_i}=1$). The averages are computed running $100$ independent simulations.}
\label{fig:blog net}
\end{figure}

\begin{figure}
\centering
\includegraphics[width = 0.7\textwidth]{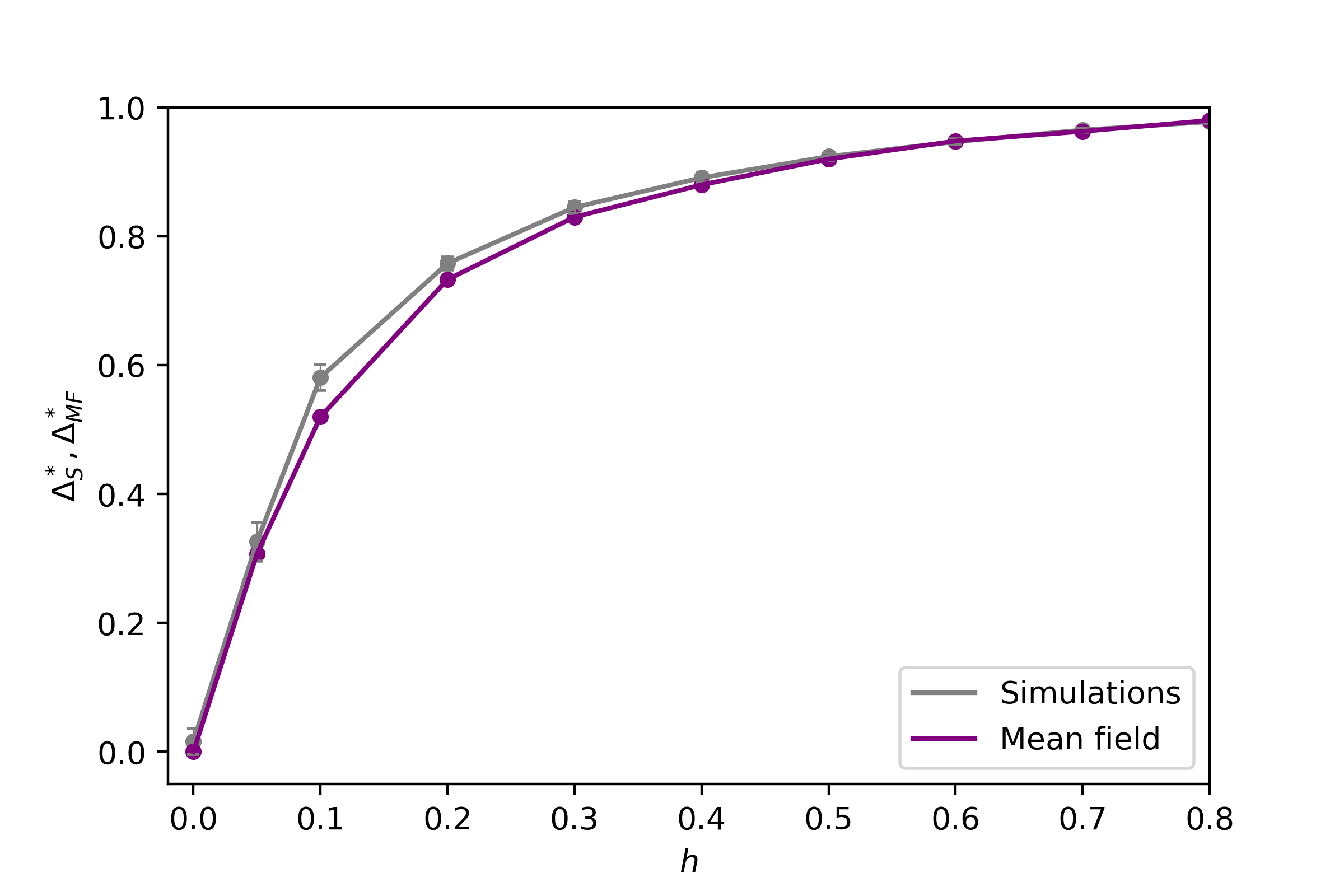}
\caption{\textbf{Mean-field predictions and simulations of the model on the 2004 Blogosphere.} We compare the mean-field predictions and the numerical simulations of the asymptotic polarization on the 2004 Political Blogosphere, restricting to equally strong biases $h_r = -h_b = h$, for various $h$.}
\label{fig:MF and blog net}
\end{figure}

We run the bipopulated VMP on a real social network with a moudular structure, to study how well the mean-field approximation performs on a real network with high modularity and potentially structural features as well as dynamical correlations. We take the network of Political Blogs during 2004 American elections \cite{adamic2005political}, characterized by the presence of two communities that reflect political bi-partisanships. After eliminating the nodes with degree less than $4$, and transforming for simplicity the original directed network in undirected (such that $p_{12}=p_{21}$), we apply a community detection algorithm \cite{kernighan1970efficient}. It turns out that the two communities, named "reds" and "blues" (shown in the upper plot of figure \ref{fig:blog net}), have sizes respectively $N_r = 413 $, $N_b= 490 $, average internal degrees $z_{rr} = 34.02 $, $z_{bb} = 32.38 $ and external degrees $z_{rb} = 2.88$, $z_{br} = 2.43$. We perform numerical simulations of the model choosing, for simplicity, opposite and equally strong preferences ($h_1 = -h_2 = h$), for multiple values of $h$, and compare them to the mean field predictions on a SBM with the same average degrees. The results are reported in figure \ref{fig:MF and blog net}: the mean-field predictions agree well with the empirical simulations, with a small gap emerging for weak preference intensities. In the lower plot of figure \ref{fig:blog net} it is reported the average polarization of each node in the asymptotic state ($t=50$) over repeated runs of the model with equal and opposite biases fixed to $h=0.3$: as one could expect, the most open-minded nodes adopt their unpreferred opinion more frequently than the ones in the rest of their community. Despite this effect due to the heterogeneity of the degrees and specifically of the local open-mindednesses, the mean-field predictions remain quite accurate.

\section{Conclusions and perspectives} \label{chapter_discussion}
In this work, we have studied a system of individuals whose process of  opinion formation is influenced by three factors: the imitative mechanism at the root of most of the models of opinion dynamics, the heterogeneous personal preferences of individuals for one opinion rather than the other, and the homophilic phenomenon at the root of the so-called \textit{epistemic bubbles}, which implies that each individual is more connected to individuals with the same preference and leads to the formation of a social network with a modular structure. We have considered two oppositely biased populations with preferences of different intensities, interacting through a social network that reflects the phenomenon of epistemic bubbles, where two individuals with the same bias happen more likely to get connected. We have derived the system of differential equations governing the opinion dynamics within the mean-field and pair approximations. In the mean-field framework, we have analytically determined the conditions under which the individuals of one population manage to induce the whole system to converge to their preferred opinion. Moreover, we have shown that the achievement of consensus depends mostly on the topological structure  of the "losing" population rather than on the one of the "winning" population. This disparity is more evident the greater is the bias intensity of the "winning" group with respect to the "losing" one.\\
\\
The appropriatness of the assumptions at the basis of the abstract and minimalistic Voter Model with Preferences needs to be tested in a real setting through the implementation of social experiments \cite{peralta2022opinion,centola2015spontaneous}. 
On the other hand, in perspective of an application of the VMP on real data, one has to cope with the problem of identifying and quantifying the external preferences attached to individuals. This is of course an open and difficult problem: a practical method would be to analyze historical positions of each individual on other topics (for example, in the context of misinformation \cite{masuda2011can}, the attached bias may correspond to the frequency at which conspiracy theories has been preferred to mainstream news in the past, by the individual). Another approach would be to first identify communities in a social network of interactions and then infer the average preferences of the individuals of the communities by analyzing the opinion uploads during internal and external interactions (similarly to \cite{pansanella2022change})\footnote{Such inverse problem has been extensively studied with mathematical rigour for the multipopulated Ising-like models on the fully connected graph \cite{fedele2013inverse,contucci2022inverse} and applied \cite{burioni2015enhancing}.}.\\
\\
Additionally, the bipopulated VMP presented in this work serves as a foundation for more sophisticated and realistic models. For example, preferences can be formulated to depend on the current opinions of the two groups. Moreover, the assumption of assigning the same preference to all individuals in a community can be relaxed to explore the impact of different distributions on the asymptotic state of the system. Furthermore, one can consider the presence of noise or other biases (e.g. the algorithmic one \cite{peralta2021effect}) or consider comparable time-scales for personal preferences and opinion uploads.\\
\\
Finally, the VMP can be compared to other models with similar settings (e.g. \cite{malarz2006truth,hernandez2013heterogeneous}), to study the roles of preferences and homophily in the opinion formation process through mutiple perspectives and approaches, as suggested by the authors of \cite{flache2017models}.
\section*{Acknowledgement} 
The research reported in this work was partially supported by the EU H2020 ICT48 project "Humane AI Net" under contract \# 952026, by the European Union – Horizon 2020 Program under the scheme “INFRAIA-01-2018-2019 – Integrating Activities for Advanced Communities”, Grant Agreement n.871042, “SoBigData++: European Integrated Infrastructure for Social Mining and Big Data Analytics”, and by the CHIST-ERA grant "SAI": CHIST-ERA-19-XAI-010, FWF (grant No. I 5205). The authors thank Antonio F. Peralta for useful discussions.

\renewcommand*{\bibfont}{\small}
\printbibliography

\appendix
\section{Linear stability analysis}
\subsection{Fully connected network}
Here we perform the linear stability analysis of the dynamical system (\ref{fc mf rho12}) derived for the VMP on the fully connected topology, giving a proof to the considerations in chapter \ref{FC_chapter}. As said in the main text, the consensus points $(0,0)$ and $(\alpha,1-\alpha)$ are fixed points of the system for whatever choice of the parameters $\alpha, h_1,h_2$, while it is easy to prove that only for $\alpha\in(\alpha_c^{-},\alpha_c^{+})$ the fixed point corresponding to the \textit{polarized} state (\ref{pol_rho},\ref{pol_delta}) is in the rectangle $(0,\alpha)\times(0,1-\alpha)$, and thus has a physical meaning. In the following, we prove that for conflicting preferences, i.e. $h_1\geq0, h_2\leq0$, for $\alpha<\alpha_c^{-}$ the negative consensus fixed point is the only stable fixed point of the system, for $\alpha\in(\alpha_c^{-},\alpha_c^{+})$ both the consensus points are unstable and the polarized fixed point is stable, while for $\alpha>\alpha_c^{-}$ the positive consensus is the only stable fixed point.. \\
\\
The Jacobian of the dynamical system reads
\begin{equation}
    J(\rho_1,\rho_2;\alpha,h_1,h_2) = \frac{1}{2}
    \begin{pmatrix}
    \alpha-1+h_1(1+\alpha-4\rho_1-2\rho_2) & \alpha+h_1(\alpha-2\rho_1)  \\
    1-\alpha +h_2(1-\alpha-2\rho_2) & -\alpha +h_2(2-\alpha-4\rho_2-2\rho_1) \\
    \end{pmatrix}
\end{equation}
According to the linear stability theory, a fixed point is stable if both the eigenvalues of the corresponding Jacobian are negative, i.e. if the trace $T$ is negative and the determinant $D$ is positive.\\
\\
For the negative consensus fixed point we have that 
\begin{align}
    & T = \frac{1}{2}\bigg[\alpha(h_1-h_2)+2h_2+h_1-1\bigg] \\
    &D = \frac{1}{2}\bigg[h_1h_2-h_2-\alpha(h_1-h_2)\bigg] 
\end{align}
so the determinant is positive for $\alpha<\alpha^D=\frac{-h_2(1-h_1)}{h1-h_2}$ , while the trace is negative for $\alpha<\alpha^T=\frac{1-2h_2-h_1}{h_1-h_2}$. Thus the stability condition is satisfied for $\alpha<\mbox{min}(\alpha^T,\alpha^D)$. It is easy to see that $\alpha^D\leq\alpha^T$, since for the assumptions on the sign of the preferences it holds that $h_1h_2 + h_1 +h_2 \leq1$, so the negative consensus point is a stable attractive fixed point if and only if
$\alpha<\alpha^D = \alpha_c^{-}$.\\
\\
Analogous considerations apply to the positive consensus fixed points, whose determinant and trace read
\begin{align}
    & T = \frac{1}{2}\bigg[-\alpha(h_1-h_2)-h_1-2h_2-1\bigg] \\
    & D = \frac{1}{2}\bigg[\alpha(h_1-h_2) + h_1h_2 + h_2 \bigg] 
\end{align}
and by applying the same arguments as before we get that the stability condition is fulfilled when $\alpha>\mbox{max}(\alpha^D,\alpha^T) = \alpha^D = \alpha_c^{+}$.\\
\\
Last, the same considerations about the trace and determinant can be used to prove that in the range $[\alpha_c^{-},\alpha_c^{+}]$ the fixed point corresponding to the \textit{polarized} state is stable.

\subsection{Modular network} \label{Modular lsa}
We perform the linear stability analysis of the mean-field system (\ref{mf_echochambers}) derived from the VMP on a modular network with two communities and open-mindedness parameters $\gamma_1,\gamma_2$. The dynamical variables taken into consideration are now $\rho'_1,\rho'_2$ as defined in chapter \ref{chapter SBM}, both in the range $[0,1]$. The analysis focuses on the stability of the consensus points, now $(0,0)$ for the negative and $(1,1)$ for the positive, that are fixed points of the system for whatever choice of the parameters. \\
\\
The Jacobian matrix of the system (\ref{mf_echochambers}), discarding the uninfluential factor, reads
\begin{equation}
\small    J(\rho'_1,\rho'_2) = 
     \begin{pmatrix}
        \gamma_1(4h_1\rho_1-2h_1\rho_2-h_1-1)-4h_1\rho_1+2h_1 & \gamma_1[1-h_1(2\rho_1-1)]\\
        \gamma_2[1-h_2(2\rho_2-1)]  &  \gamma_2(4h_2\rho_2-2h_2\rho_1-h_2-1)-4h_2\rho_2+2h_2 \\
    \end{pmatrix}
\end{equation}
and as before we compute the traces and the determinant of the Jacobian at the fixed point in order to obtain the stability conditions. \\
\\
The traces and determinant for the positive consensus point read
\begin{align}
    & T = -\gamma_1(1-h_1)-\gamma_2(1-h_2)-2(h_1+h_2)\\
    & D = 2\big[ \gamma_1h_2(1-h_1) +\gamma_2h_1(1-h_2)+2h_1h_2 \big]
\end{align}
and we see that the trace is negative for all $h_1\geq -h_2$, while the determinant is positive for 
\begin{equation}
    \gamma_1h_2(1-h_1)+\gamma_2h_1(1-h_2) + 2h_1h_2>0
\label{det_condition}
\end{equation}
However, it never happens that $T>0$ and $D>0$ at the same time, at least in the parameters' space of interest $h_1\geq0,h_2\leq0,\gamma_{1/2}\geq0$. To prove it, we try to solve the system
\begin{equation}
\begin{cases}
    T = -\gamma_1(1-h_1)-\gamma_2(1+|h_2|)-2(h_1-|h_2|)>0\\
    D \propto -\gamma_1|h_2|(1-h_1) +\gamma_2h_1(1+|h_2|)-2h_1|h_2|>0
\end{cases}
\end{equation}
Arranging the terms we are left with the series of inequalities
\begin{equation}
    -\gamma_1(1-h_1)-2(h_1-|h_2|) > \gamma_2(1+|h_2|) > \frac{1}{h_1} \big(\gamma_1|h_2|(1-h_1) +2h_1|h_2|\big)
\end{equation}
that implies 
\begin{equation}
    -\gamma_1(1-h_1)-2(h_1-|h_2|)  > \frac{1}{h_1} \big(\gamma_1|h_2|(1-h_1) +2h_1|h_2|\big)
\end{equation}
and simplifies in 
\begin{equation}
       -\gamma_1h_1(1-h_1)-2h_1^2 >  \gamma_1|h_2|(1-h_1)
\end{equation}
which is never true, since the terms on the l.h.s. are always negative and the term on the r.h.s. is positive. Thus, the whole stability region of the positive consensus fixed point is determined by the condition derived from the determinant (\ref{det_condition}), and thus delimited by the critical curve
\begin{equation}
   \gamma_1h_2(1-h_1) +\gamma_2h_1(1-h_2)+2h_1h_2 = 0
\end{equation}
\\
For the negative consensus we have 
\begin{align}
    & T = -\gamma_1(1+h_1)-\gamma_2(1+h_2)+2(h_1+h_2)\\
    & D = 2\big[ -\gamma_1h_2(1+h_1) -\gamma_2h_1(1+h_2)+2h_1h_2  \big]
\end{align}
and applying the same arguments of the positive consensus we can claim that the stability condition is determined only by the condition on the determinant $D>0$, thus the corresponding critical curve reads
\begin{equation}
    -\gamma_1h_2(1+h_1) -\gamma_2h_1(1+h_2)+2h_1h_2 = 0
\end{equation}

\section{Mean-field transition rates for the modular network}\label{app transition rates}
As in the fully connected case, each of the global mean-field transition rates (\ref{SBM transition rates}) for the modular network is the product of three factors: the probability to randomly select an agent of class $i$ and current state $\sigma$, the probability of selecting one neighbour of such agent type currently in the opposite state $-\sigma$, and the probability of transition (imitation). With respect to the fully connected case, the first and the third factor are obviously unchanged, and in the case of $R_{+1}$ correspond respectively to $\alpha-\rho_1$ and $\frac{1+h_1}{2}$. To determine the second factor, we have to take carefully into account the modular structure and distinguish the two classes. For $R_{+1}$, once selected a spin of of the first class, the probability of randomly selecting a neighbouring agent of the first class is $\frac{\alpha p_{11}}{\alpha p_{11} + (1-\alpha)p_{12}}$, multiplied by the probability that such neighbour is in the up state $\frac{\rho_1}{\alpha}$. Analogously, the probability of randomly selecting a neighbour of the second class is $\frac{(1-\alpha)p_{12}}{\alpha p_{11} + (1-\alpha)p_{12}}$, multiplied by the probability that such neighbour is in the up state $\frac{\rho_2}{1-\alpha}$. The result is the factor  $\frac{p_{11}\rho_1}{\alpha p_{11} + (1-\alpha)p_{12}}+\frac{p_{12}\rho_2}{\alpha p_{11} + (1-\alpha)p_{12}}$. Analogous considerations apply for the other transition rates $R_{1-},R_{2+}$ and $R_{2-}$.

\section{Derivation of the Pair Approximation system of ODEs}\label{app_PA}
For the sake of simplicity, we apply the pair approximation on a z-regular undirected modular graph with two communities, but the treatment can be easily extended to a directed modular network with heterogeneous degrees.\\
Each node of class $1$ has $z_{11}$ internal neighbours and $z_{12}$ external (of different class) ones. Same thing for class 2. Moreover, $N_1z_{12} = N_2 z_{21}$. The following quantities are defined: the density of up spins in the first and second class, respectively $\rho'_1$ and $\rho'_2,$ as in the paragraph before, while $b_{lm}^{\sigma_i\;\sigma_j}$ represents the fraction of active\footnote{An edge linking the nodes $ij$ is active at a time $t$ if $\sigma_i(t)\sigma_j(t)=-1$} links connecting a spin in state $\sigma_i$ spin of class $l$ with a spin in state $\sigma_j$ spin of class $m$, normalized by the total number of connections of the class $l$ towards the class $m$,
\begin{align}
    &\rho'_1 = \frac{1}{N_1}\sum_{i=1}^{N_1} \frac{1+\sigma_i}{2} \\
    &\rho'_2 = \frac{1}{N_2}\sum_{i=N_1+1}^{N} \frac{1+\sigma_i}{2} \\
    &b_{11}^{+-} = \frac{1}{N_1z_{11}}\sum_{i=1}^{N_1}\;\sum_{j=1}^{N_1}A_{ij}\frac{1-\sigma_i+\sigma_j-\sigma_i\sigma_j}{4}  \\
    &b_{22}^{+-} = \frac{1}{N_2z_{22}}\sum_{i=N_1+1}^{N}\;\sum_{j=N_1+1}^{N}A_{ij}\frac{1-\sigma_i+\sigma_j-\sigma_i\sigma_j}{4}  \\
    &b_{12}^{-+} = \frac{1}{N_1z_{12}}\sum_{i=1}^{N_1}\sum_{j=N_1+1}^{N}A_{ij}\frac{1-\sigma_i+\sigma_j-\sigma_i\sigma_j}{4}  \\
    &b_{12}^{+-} = \frac{1}{N_1z_{12}}\sum_{i=1}^{N_1}\sum_{j=N_1+1}^{N}A_{ij}\frac{1+\sigma_i-\sigma_j-\sigma_i\sigma_j}{4}
\end{align}
Due to the undirectedness of the network, the other quantities of interest can be expressed as functions of the ones defined above, specifically $b_{11}^{-+} = b_{11}^{+-}$, $b_{22}^{-+} = b_{22}^{+-}$, $b_{21}^{+-} = b_{12}^{-+}$, $b_{21}^{-+} = b_{12}^{+-}$.\\
\\
The first of the global rates of the possible processes $W^{1+}_{z_{11},m_{11},z_{12},m_{12}}$, i.e. the probability that in a time $dt$ (recall $dt=\frac{1}{N}d\tau$) a spin of the first class in current state $-$ and with $m_{11}$ out of $z_{11}$ internal neighbours and $m_{12}$ out of $z_{12}$ external ones currently in $+$ state flips, reads
\begin{align}
    &W^{1-\rightarrow+}_{z_{11},m_{11},z_{12},m_{12}} = N_1 (1-\rho'_1)P^{11-}_{z_{11},m_{11}}P^{12-}_{z_{12},m_{12}}F^1_{z_{11}+z_{12},m_{11}+m_{12}}
\end{align}
where $P^{11-}_{z_{11},m_{11}}$ is the probability that a node in the first population currently in state $-1$ has $z_{11}$  degree (trivial, explicited only for the generalization) and $m_{11}$ neighbours of the first population currently in state $+1$. The other rates read similarly
\begin{align}
    &W^{2-\rightarrow+}_{z_{22},m_{22},z_{21},m_{21}} = N_2 (1-\rho'_2)P^{22-}_{z_{22},m_{22}}P^{21-}_{z_{21},m_{21}}F^2_{z_{22}+z_{21},m_{22}+m_{21}}\\
    &W^{1+\rightarrow-}_{z_{11},m_{11},z_{12},m_{12}} = N_1 \rho'_1P^{11+}_{z_{11},m_{11}}P^{12+}_{z_{12},m_{12}}R^1_{z_{11}+z_{12},m_{11}+m_{12}}\\
    &W^{2+\rightarrow-}_{z_{22},m_{22},z_{21},m_{21}} = N_2 \rho'_2P^{22+}_{z_{22},m_{22}}P^{21+}_{z_{21},m_{21}}R^2_{z_{22}+z_{21},m_{22}+m_{21}}
\end{align}
The transition rates (infection $-1\rightarrow +1$ and recovery $+1\rightarrow -1$) in the expressions of the global rates, for this specific multi-class binary-state stochastic process, are 
\begin{align}
    &F^{1/2}_{z,m} = \frac{m}{z}\frac{1+h_{1/2}}{2}\\
    &R^{1/2}_{z,m} = \bigg(1-\frac{m}{z}\bigg)\frac{1-h_{1/2}}{2}
\end{align}
The two global rates of the first class $W^{1\mp\rightarrow\pm}_{z_{11},m_{11},z_{12},m_{12}}$ change the variables respectively
\begin{align}
    &\rho'_1 \rightarrow \rho'_1 \pm \frac{1}{N_1}\\
    &b_{11}^{+-} \rightarrow b_{11} \pm \frac{(z_{11}-2m_{11})}{N_1z_{11}}  \\
    &b_{12}^{-+} \rightarrow  b_{12}^{-+} \mp \frac{m_{12}}{N_1z_{12}} \\
    &b_{12}^{+-} \rightarrow  b_{12}^{+-} \pm \frac{z_{12}-m_{12}}{N_1z_{12}}
\end{align}
and correspondingly the ones the second class rates $W^{2\mp\rightarrow\pm}_{z_{22},m_{22},z_{21},m_{21}}$ change the variables 
\begin{align}
    &\rho'_2 \rightarrow \rho'_2 \pm \frac{1}{N_2}\\
    &b_{22}^{+-} \rightarrow b_{22} \pm \frac{(z_{22}-2m_{22})}{N_2z_{22}}  \\
    &b_{12}^{-+} \rightarrow  b_{12}^{-+} \pm  \frac{z_{21}-m_{21}}{N_2z_{21}}\\
    &b_{12}^{+-} \rightarrow  b_{12}^{+-} \mp  \frac{m_{21}}{N_2z_{21}} 
\end{align}
\\
Thus, the dynamical system consists of six coupled evolution equations 
\begin{equation}
    \begin{cases}
    \dot{\rho'_1} = \frac{1}{N_1} \sum\limits_{m_{11}=0}^{z_{11}}\sum\limits_{m_{12}=0}^{z_{12}}[W^{1+}_{z_{11},m_{11},z_{12},m_{12}} - W^{1-}_{z_{11},m_{11},z_{12},m_{12}}]\\
    \dot{\rho'_2} = \frac{1}{N_2} \sum\limits_{m_{22}=0}^{z_{22}}\sum\limits_{m_{21}=0}^{z_{21}}[W^{2+}_{z_{22},m_{22},z_{21},m_{21}} - W^{2-}_{z_{22},m_{22},z_{21},m_{21}}]\\
    \dot{b_{11}^{+-}} = \frac{1}{N_1z_{11}} \sum\limits_{m_{11}=0}^{z_{11}}\sum\limits_{m_{12}=0}^{z_{12}}(z_{11} - 2m_{11})[W^{1+}_{z_{11},m_{11},z_{12},m_{12}} - W^{1-}_{z_{11},m_{11},z_{12},m_{12}}]\\
    \dot{b_{22}^{+-}} = \frac{1}{N_2z_{22}} \sum\limits_{m_{22}=0}^{z_{22}}\sum\limits_{m_{21}=0}^{z_{21}}(z_{22} - 2m_{22})[W^{2+}_{z_{22},m_{22},z_{21},m_{21}} - W^{2-}_{z_{22},m_{22},z_{21},m_{21}}]\\
    \dot{b_{12}^{-+}} = \frac{1}{N_1z_{12}} \sum\limits_{m_{11}=0}^{z_{11}}\sum\limits_{m_{12}=0}^{z_{12}}(-m_{12})[W^{1+}_{z_{11},m_{11},z_{12},m_{12}} - W^{1-}_{z_{11},m_{11},z_{12},m_{12}}]  \;+\\ 
    \;\;\;\;\;\;\;\;\;\;\;\;\;+\frac{1}{N_2z_{21}} \sum\limits_{m_{22}=0}^{z_{22}}\sum\limits_{m_{21}=0}^{z_{21}}(z_{21} - m_{21})[W^{2+}_{z_{22},m_{22},z_{21},m_{21}} - W^{2-}_{z_{22},m_{22},z_{21},m_{21}}]\\
    \dot{b_{12}^{+-}} = \frac{1}{N_1z_{12}} \sum\limits_{m_{11}=0}^{z_{11}}\sum\limits_{m_{12}=0}^{z_{12}}(z_{12}-m_{12})[W^{1+}_{z_{11},m_{11},z_{12},m_{12}} - W^{1-}_{z_{11},m_{11},z_{12},m_{12}}]  \;+\\ 
    \;\;\;\;\;\;\;\;\;\;\;\;\;+\frac{1}{N_2z_{21}} \sum\limits_{m_{22}=0}^{z_{22}}\sum\limits_{m_{21}=0}^{z_{21}}( - m_{21})[W^{2+}_{z_{22},m_{22},z_{21},m_{21}} - W^{2-}_{z_{22},m_{22},z_{21},m_{21}}]\\    
    
    \end{cases}
\end{equation}
We still have to defined the probabilities e.g.  $P^{11-}_{z_{11},m_{11}}$, whose definition would close the system of ordinary differential equations above.\\
\\
The mean-field approximation corresponds to take the probabilities 
\begin{align}
    &P^{11+}_{z_{11},m_{11}} = P^{11-}_{z_{11},m_{11}} =  \delta(\frac{m_{11}}{z_{11}} - \rho'_1) \label{MF_probabilities_start} \\
    &P^{22+}_{z_{22},m_{22}} = P^{22-}_{z_{22},m_{22}} =  \delta(\frac{m_{22}}{z_{22}}-\rho'_2)\\
    &P^{12+}_{z_{12},m_{12}} = P^{12-}_{z_{12},m_{12}} = \delta(\frac{m_{12}}{z_{12}}-\rho'_2) \\
    &P^{21+}_{z_{21},m_{21}} = P^{21-}_{z_{21},m_{21}} = \delta(\frac{m_{21}}{z_{21}}-\rho'_1)
\label{MF_probabilities_end}
\end{align}
and thus decouples the evolution of the densities of up spin from the rest of the system, indeed by inserting those probabilities we would recover the mean-field evolution equations (\ref{mf_echochambers}) for $\rho'_1,\rho'_2$ .\\
\\
The pair approximation, instead, consists in assuming that those probability are binomial distributions $B_{z,m}(x) = \binom{z}{m}x^m(1-x)^{z-m}$ 
\begin{align} 
    &P^{11-}_{z_{11},m_{11}} = B_{z_{11},m_{11}}(p_{11-}) \label{eq:PA_proba_start}\\
    &P^{11+}_{z_{11},m_{11}} = B_{z_{11},z_{11}-m_{11}}(p_{11+}) \\
    &P^{22-}_{z_{22},m_{22}} = B_{z_{22},m_{22}}(p_{22-}) \\
    &P^{22+}_{z_{22},m_{22}} = B_{z_{22},z_{22}-m_{22}}(p_{22+}) \\
    &P^{12-}_{z_{12},m_{12}} = B_{z_{12},m_{12}}(p_{12-}) \\
    &P^{12+}_{z_{12},m_{12}} = B_{z_{12},z_{12}-m_{12}}(p_{12+}) \\
    &P^{21-}_{z_{21},m_{21}} = B_{z_{21},m_{21}}(p_{21-}) \\
    &P^{21+}_{z_{21},m_{21}} = B_{z_{21},z_{21}-m_{21}}(p_{21+}) \label{PA_proba_end}
\end{align}
with single events probabilities
\begin{align}
    &p_{11-} = \frac{b_{11}^{-+}}{1-\rho'_1} = \frac{b_{11}^{+-}}{1-\rho'_1}\label{PA_single_proba_start}\\
    &p_{11+} = \frac{b_{11}^{+-}}{\rho'_1} \\
    &p_{22-} = \frac{b_{22}^{-+}}{1-\rho'_2} = \frac{b_{22}^{+-}}{1-\rho'_2} \\
    &p_{22+} = \frac{b_{22}^{+-}}{\rho'_2} \\
    &p_{12-} = \frac{b_{12}^{-+}}{1-\rho'_1} \\
    &p_{12+} = \frac{b_{12}^{+-}}{\rho'_1} \\
    &p_{21-} = \frac{b_{21}^{+-}}{1-\rho'_2} = \frac{b_{12}^{-+}}{1-\rho'_2}  \\
    &p_{21+} =  \frac{b_{21}^{+-}}{\rho'_2} = \frac{b_{12}^{-+}}{\rho'_2}\label{PA_single_proba_end}   
\end{align}
The criterium is to consider as single probability, e.g. to express $p_{11-}$, the fraction of $-+$ links of the first communities (in number $b_{11}^{-+}N_1z_{11}$) over the total of the edges starting from $-$ within the first community (in number $(1-\rho'_1)N_1z_{11}$).\\
\\
Eventually we can write the final system of ODE within the pair approximation

\begin{equation}
    \begin{cases}
    \dot{\rho'_1} =   (1-\rho'_1)\sum\limits_{m_{11}=0}^{z_{11}}B_{z_{11},m_{11}}(p_{11-})\sum\limits_{m_{12}=0}^{z_{12}}B_{z_{12},m_{12}}(p_{12-})F^1_{z_{11}+z_{12},m_{11}+m_{12}} +\\
    \;\;\;\;\;\;\;\;\; -  \rho'_1 \sum\limits_{m_{11}=0}^{z_{11}}B_{z_{11},z_{11}-m_{11}}(p_{11+}) \sum\limits_{m_{12}=0}^{z_{12}}B_{z_{12},z_{12}-m_{12}}(p_{12+})R^1_{z_{11}+z_{12},m_{11}+m_{12}}\\
    \\
    
    \dot{\rho'_2} = (1-\rho'_2)\sum\limits_{m_{22}=0}^{z_{22}}B_{z_{22},m_{22}}(p_{22-}) \sum\limits_{m_{21}=0}^{z_{21}}B_{z_{21},m_{21}}(p_{21-})F^2_{z_{22}+z_{21},m_{22}+m_{21}} +\\
    \;\;\;\;\;\;\;\;\;- \rho'_2\sum\limits_{m_{22}=0}^{z_{22}}B_{z_{22},z_{22}-m_{22}}(p_{22+})\sum\limits_{m_{21}=0}^{z_{21}}B_{z_{21},z_{21}-m_{21}}(p_{21+})R^2_{z_{22}+z_{21},m_{22}+m_{21}}\\
    \\
    
    \dot{b_{11}^{+-}} = \frac{1-\rho'_1}{z_{11}} \sum\limits_{m_{11}=0}^{z_{11}}\sum\limits_{m_{12}=0}^{z_{12}}(z_{11} - 2m_{11})B_{z_{11},m_{11}}(p_{11-})B_{z_{12},m_{12}}(p_{12-})F^1_{z_{11}+z_{12},m_{11}+m_{12}} +\\
    \;\;\;\;\;\;\;\;\;\;-\frac{\rho'_1}{z_{11}} \sum\limits_{m_{11}=0}^{z_{11}}\sum\limits_{m_{12}=0}^{z_{12}}(z_{11} - 2m_{11})B_{z_{11},z_{11}-m_{11}}(p_{11+}) B_{z_{12},z_{12}-m_{12}}(p_{12+})R^1_{z_{11}+z_{12},m_{11}+m_{12}} \\
    \\
    
    \dot{b_{22}^{+-}} = \frac{1-\rho'_2}{z_{22}} \sum\limits_{m_{22}=0}^{z_{22}}\sum\limits_{m_{21}=0}^{z_{21}}(z_{22} - 2m_{22})B_{z_{22},m_{22}}(p_{22-}) B_{z_{21},m_{21}}(p_{21-})F^2_{z_{22}+z_{21},m_{22}+m_{21}} +\\ \;\;\;\;\;\;\;\;\;-\frac{\rho'_2}{z_{22}}\sum\limits_{m_{22}=0}^{z_{22}}\sum\limits_{m_{21}=0}^{z_{21}}(z_{22} - 2m_{22})B_{z_{22},z_{22}-m_{22}}(p_{22+})B_{z_{21},z_{21}-m_{21}}(p_{21+})R^2_{z_{22}+z_{21},m_{22}+m_{21}}\\
    \\
    
    \dot{b_{12}^{-+}} = \frac{1-\rho'_1}{z_{12}} \sum\limits_{m_{11}=0}^{z_{11}}\sum\limits_{m_{12}=0}^{z_{12}}(-m_{12})B_{z_{11},m_{11}}(p_{11-})B_{z_{12},m_{12}}(p_{12-})F^1_{z_{11}+z_{12},m_{11}+m_{12}} + \\ \;\;\;\;\;\;\;\;\;\;\;-\frac{\rho'_1}{z_{12}}\sum\limits_{m_{11}=0}^{z_{11}}\sum\limits_{m_{12}=0}^{z_{12}}(-m_{12}) B_{z_{11},z_{11}-m_{11}}(p_{11+}) B_{z_{12},z_{12}-m_{12}}(p_{12+})R^1_{z_{11}+z_{12},m_{11}+m_{12}}  \;+\\ 
    \;\;\;\;\;\;\;\;\;\;\;\;\;+\frac{1-\rho'_2}{z_{21}} \sum\limits_{m_{22}=0}^{z_{22}}\sum\limits_{m_{21}=0}^{z_{21}}(z_{21} - m_{21})B_{z_{22},m_{22}}(p_{22-}) B_{z_{21},m_{21}}(p_{21-})F^2_{z_{22}+z_{21},m_{22}+m_{21}}\;+\\
    \;\;\;\;\;\;\;\;\;\;\;\;- \frac{\rho'_2}{z_{21}}\sum\limits_{m_{22}=0}^{z_{22}}\sum\limits_{m_{21}=0}^{z_{21}}(z_{21} - m_{21})B_{z_{22},z_{22}-m_{22}}(p_{22+})B_{z_{21},z_{21}-m_{21}}(p_{21+})R^2_{z_{22}+z_{21},m_{22}+m_{21}}\\
    \\
    
    \dot{b_{12}^{+-}} = \frac{1-\rho'_1}{z_{12}} \sum\limits_{m_{11}=0}^{z_{11}}\sum\limits_{m_{12}=0}^{z_{12}}(z_{12}-m_{12})B_{z_{11},m_{11}}(p_{11-})B_{z_{12},m_{12}}(p_{12-})F^1_{z_{11}+z_{12},m_{11}+m_{12}} +\\
    \;\;\;\;\;\;\;\;\;\;\;\;-  \frac{\rho'_1}{z_{12}}\sum\limits_{m_{11}=0}^{z_{11}}\sum\limits_{m_{12}=0}^{z_{12}}(z_{12}-m_{12}) B_{z_{11},z_{11}-m_{11}}(p_{11+}) B_{z_{12},z_{12}-m_{12}}(p_{12+})R^1_{z_{11}+z_{12},m_{11}+m_{12}}   \;+\\ 
    \;\;\;\;\;\;\;\;\;\;\;\;\;+\frac{1-\rho'_2}{z_{21}} \sum\limits_{m_{22}=0}^{z_{22}}\sum\limits_{m_{21}=0}^{z_{21}}( - m_{21})B_{z_{22},m_{22}}(p_{22-}) B_{z_{21},m_{21}}(p_{21-})F^2_{z_{22}+z_{21},m_{22}+m_{21}} +\\
    \;\;\;\;\;\;\;\;\;\;\;\;- \frac{\rho'_2}{z_{21}}\sum\limits_{m_{22}=0}^{z_{22}}\sum\limits_{m_{21}=0}^{z_{21}}( - m_{21})B_{z_{22},z_{22}-m_{22}}(p_{22+})B_{z_{21},z_{21}-m_{21}}(p_{21+})R^2_{z_{22}+z_{21},m_{22}+m_{21}}\\    
    \end{cases}
    \label{PA_system}
\end{equation}
to be solve numerically with standard methods.\\
From the initial conditions $\rho'_1(0),\rho'_2(0)$, the other initial conditions are determined as follows
\begin{align}
    & b_{11}^{+-}(0) = \rho'_1(0)(1-\rho'_1(0))\\
    & b_{22}^{+-}(0) = \rho'_2(0)(1-\rho'_2(0))\\
    & b_{12}^{-+}(0) = \rho'_1(0)(1-\rho'_2(0))\\
    & b_{12}^{+-}(0) = \rho'_2(0)(1-\rho'_1(0))
\end{align}

\end{document}